\documentclass[reprint,aps,prb,superscriptaddress,nobibnotes]{revtex4-1}
\usepackage{amsmath}
\usepackage{amssymb}
\usepackage{natbib}
\usepackage{graphicx}
\usepackage{color}
\usepackage{colordvi}
\usepackage{bm}
\usepackage{calrsfs}
\usepackage[thinlines,thiklines]{easybmat}

\setcitestyle{numbers,square}
\begin{document}

\title{Chiral excitation of spin waves in ferromagnetic films}
\author{Tao Yu}
\affiliation{Kavli Institute of NanoScience, Delft University of Technology,
	2628 CJ Delft, The Netherlands}
\author{Chuanpu Liu}
\affiliation{Fert Beijing Institute, BDBC, School of Microelectronics, Beihang University, Beijing 100191, China}
\author{Haiming Yu}
\affiliation{Fert Beijing Institute, BDBC, School of Microelectronics, Beihang University, Beijing 100191, China}
\author{Yaroslav M. Blanter}
\affiliation{Kavli Institute of NanoScience, Delft University of Technology,
	2628 CJ Delft, The Netherlands}
\author{Gerrit E. W. Bauer}
\affiliation{Institute for Materials Research $\&$ WPI-AIMR $\&$ CSRN,
	Tohoku University, Sendai 980-8577, Japan} \affiliation{Kavli Institute of
	NanoScience, Delft University of Technology, 2628 CJ Delft, The Netherlands}
\date{\today}

\begin{abstract}
We theoretically investigate the interlayer dipolar and exchange couplings
for an array of metallic magnetic nanowires grown on top of an extended
ultrathin yttrium iron garnet film. The calculated interlayer dipolar
coupling agrees with observed anticrossings [Chen \emph{et al.}, Phys. Rev.
Lett. \textbf{120}, 217202 (2018)], concluding that the interlayer exchange
coupling is suppressed by a spacer layer between the nanowires and film. The
Kittel mode in the nanowire array couples chirally to spin waves in the
film, even though Damon-Eshbach surface modes do not exist. The chirality is
suppressed when the interlayer exchange coupling becomes strong.
\end{abstract}

\pacs{75.75.-c,75.78.-n,75.30.Ds}
\maketitle


\section{Introduction}

Magnon spintronics is the research field aimed at understanding and
controlling spin waves --- the collective excitations of magnetic order ---
and its quanta, magnons, with perspectives of technological applications 
\cite{magnonics1,magnonics2,magnonics3,magnonics4}. Yttrium iron garnet
(YIG), a ferrimagnetic insulator, is currently the best material for magnon
spintronics due to its record low damping \cite{YIG1,YIG2,YIG3}.
Long-wavelength spin waves in YIG can travel over centimeters \cite%
{centimeter}. Dipolar interactions add unique features to the magnetostatic
surface or Damon-Eshbach (DE) spin waves in magnetic film with in-plane
magnetization that are exponentially localized at the surface and possess
directional chirality: the surface spin waves propagate only in one
direction that is governed by surface normal and magnetization directions 
\cite%
{Walker,DE,spin_waves_book,new_book,Jilei1,Jilei2,Jilei3,Jilei4,Kostylev,Jilei7}%
. This chirality can be very attractive for application in magnetic logics 
\cite{magnetic_logic}. However, dipolar surface spin waves suffer from a low
group velocity which makes them less attractive for information transfer. A
different mechanism --- exchange interactions --- generates spin waves with
much higher group velocity, but they are scattered easily. Transport is then
slowed down by becoming diffusive and their reach becomes limited to the
order of 10 $\mathrm{\mu }$m and the directional chirality vanishes as well.

\begin{figure}[th]
\begin{center}
{\includegraphics[width=9cm]{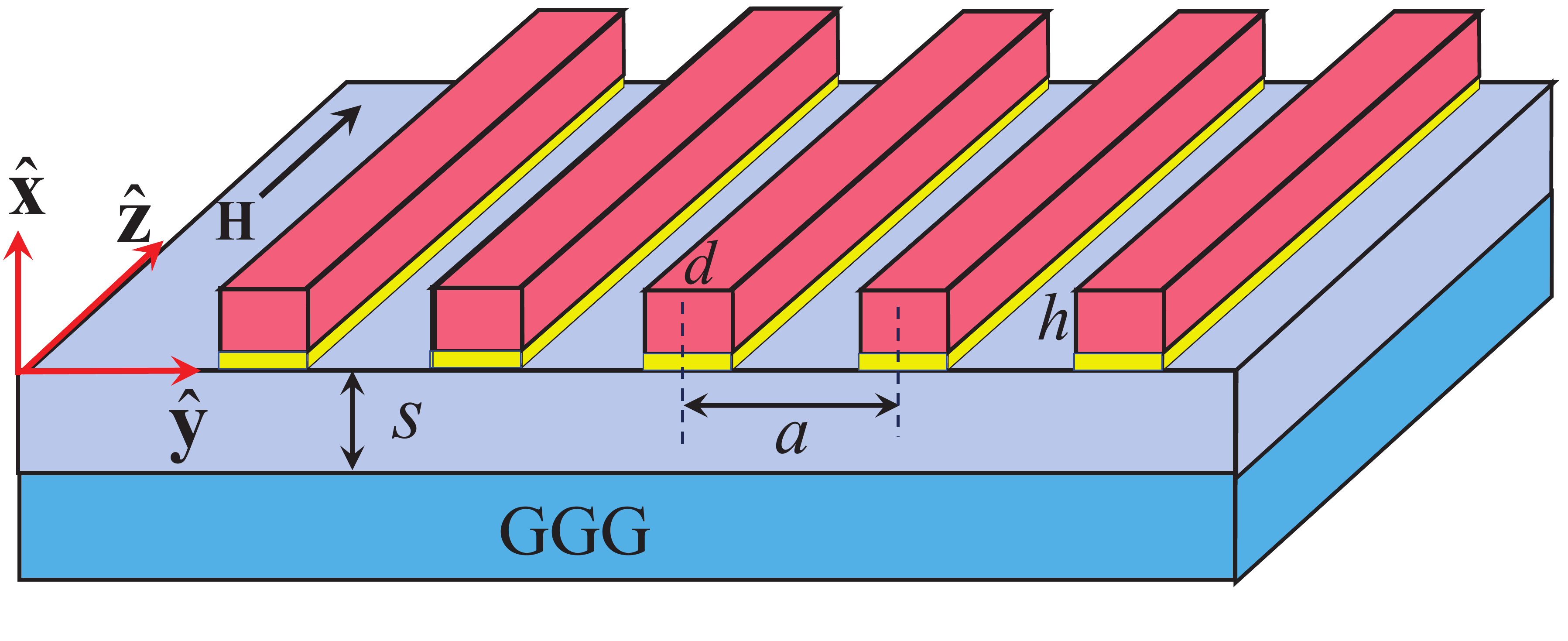}}
\end{center}
\caption{Co or Ni nanowire grating on an YIG film with coordinate system and
geometric parameters. The YIG film is fabricated on the gadolinium gallium
garnet (GGG) substrate that is a non-magnetic insulator. A magnetic field is
applied in the $\hat{\mathbf{z}}$-direction, parallel to the nanowires. A
thin non-magnetic spacer (yellow) may be inserted between the wires and the
film to suppress the interface exchange interaction.}
\label{grating1}
\end{figure}

The spin waves most suitable for information technologies therefore arise in
the intermediate regime, i.e., dipolar-exchange spin waves that combine the
long-lifetime and attractive features, such as the chirality of
magnetostatic magnons, with the higher group velocity generated when the
exchange interaction kicks in. Unfortunately, these spin waves are hard to
excite since coherent microwave absorption conserves linear momentum, and
the impedance matching problem exists when using conventional coplanar
waveguide. Recently, excitation of relatively-short--wavelength spin waves
in Co(FeB)$|$YIG thin-film bilayers with uniform microwave fields has been
demonstrated \cite{CoFeB_YIG,Co_YIG}, but these are standing waves which can
not travel. Refs.~\cite{Haiming_NC,Haiming_PRL} demonstrated that microwaves
can excite higher-momentum in-plane spin waves by ferromagnetic resonance
(FMR) of Ni or Co nanowire arrays (NWA) on an ultrathin (20 nm) YIG film
(see Fig.~\ref{grating1}). The dimensions of the grating in Fig.~\ref%
{grating1} are the thickness $h$ and width $d$ of the nanowires, the period
or center-to-center distance between the nanowires $a,$ and the YIG
thickness $s$. We choose $\hat{\mathbf{z}}$ to be parallel to the nanowires,
the magnetizations, and the applied magnetic field. A thin non-magnetic
layer between the nanowires and film suppresses the interlayer exchange
coupling. We allow NWA and YIG magnetizations to be antiparallel as well. We
investigate the magnetization dynamics of such a magnetic grating on a
magnetic film and find the spin waves can be chirally excited. This is at
the first glance surprising since DE surface modes \cite{DE} do not exist
for such thin films. However, it corresponds to and explains recent
experiments (Yu c.s., unpublished). We show that the chirality arises from
the unique polarization-momentum locking of the dipolar field generated by
the Kittel modes of NWA.

In the experiments, a coplanar waveguide on top of the NWA$|$YIG system of
Fig.~\ref{grating1} is tuned to the NWA Kittel mode, in which the
magnetization of all wires precesses in phase. Due to the large
magnetization and form anisotropy of Co and Ni, this frequency is much
higher than that of the underlying YIG film FMR. The array acts as grating
that couples to short-wavelength in-plane spin waves in the YIG film by the
dipole and exchange couplings \cite{Haiming_NC,Haiming_PRL}. Only the spin
waves propagating perpendicular to the nanowires (the $\hat{\mathbf{y}}$%
-direction in Fig.~\ref{grating1}) with in-plane wave vector $\mathbf{k}=m\pi%
\hat{\mathbf{y}}/a$ can be excited, where $m$ is an even integer. The
coherent coupling generates anti-crossings between the NWA Kittel mode and
the spin waves in the YIG film that can be observed in the microwave
reflection spectra \cite{Haiming_PRL}. The mode splitting is a direct
measure of the interlayer coupling strength. Since YIG is magnetically very
soft, the magnetizations of film and nanowires can be rotated with respect
to each other, which enhances the interlayer coupling up to GHz when in
antiparallel configuration \cite{Haiming_PRL}. We theoretically study the
dynamics of this system, focussing on the experimentally relevant
thin-YIG-film limit (e.g., $s\lesssim$20~nm). We find a good agreement with
experiments when only the dipolar coupling is taken into account, which
could indicate that the spacer in the experiments suppresses the exchange
interaction \cite{Haiming_PRL}. Interestingly, we find that the coupling is
chiral, i.e. it excites only spin waves propagating with linear momentum $%
\mathbf{k}\parallel(\mathbf{\tilde{m}}_{0}\times\mathbf{n})$, where $\mathbf{%
\tilde{m}}_{0}$ is the magnetization of and $\mathbf{n}$ the normal to the
film as is known for surface DE modes in thick films \cite{DE}. However, DE
modes do not exist in thin films with a magnetization dynamics that is
almost constant over the film thickness. The predicted chiral coupling of
exchange spin waves survives a finite interface exchange coupling and adds
functionality to down-scaled magnonic devices \cite{nanomagnetism}.

This paper is organized as follows. We first introduce the uncoupled modes
for NWA and the YIG film in Sec.~\ref{Spin_waves}. Then, the interlayer
dipolar and exchange interactions are addressed in Sec.~\ref%
{interlayer_coupling} followed by concrete calculations and comparison with
experiments in Sec.~\ref{compare_with_exp}. Finally, section~\ref{summary}
contains a discussion of the results and conclusions.

\section{Uncoupled dynamics}

\label{Spin_waves} In this section, we formulate the Kittel mode dynamics in
a NWA as well as spin waves in the thin magnetic film. The collective mode
in the NWA generates the high momentum Fourier components that couple to the
exchange spin waves in the film as elaborated in Section \ref%
{interlayer_coupling}. There we also need the spin waves amplitude
formulated in Section \ref{spin_waves_film}.

\subsection{Kittel mode in nanowire array (NWA)}

\label{Kittel_nanowires}

The NWA with a length much larger than the periodicity is to a very good
approximation a one-dimensional magnonic crystal \cite%
{magnetic_nanodots,magnonics1,crystal1,crystal2}. In this limit we may
disregard interwire dipolar interactions.

The frequency $\omega_{K}$ and magnetization amplitude $\mathbf{m}%
^{K}=\left( m_{x}^{K},m_{y}^{K}\right) $ of the Kittel mode in a single
magnetic wire read \cite{Ding_nanowire,Kittel_book} 
\begin{align}
\omega_{K} & =\mu_{0}\gamma\sqrt{(H_{\mathrm{app}}+M_{0}N_{xx} )(H_{\mathrm{%
app}}+M_{0}N_{yy})},  \label{Kittel_om} \\
m_{y}^{K} & =i\sqrt{(H_{\mathrm{app}}+M_{0}N_{xx})/(H_{\mathrm{app}}
+M_{0}N_{yy})}m_{x}^{K},  \label{Kittel_f}
\end{align}
where $\mu_{0}$ is the vacuum permeability, $-\gamma$ the electron
gyromagnetic ratio, $H_{\mathrm{app}}$ the applied magnetic field (in the $%
\hat{\mathbf{z}}$-direction), $M_{0}$ the saturation magnetization, and $%
N_{\alpha\alpha}$ the demagnetization factor with $N_{zz}=0$ for a
sufficiently long wire \cite%
{Kittel_book,Kittel_1948,Ding_nanowire,A_Aharoni,ellipsoid_1945,Haiming_NC,Ding_nanowire}%
. When $h\ll d$ (or $h\gg d$) the demagnetization factor of ellipsoids \cite%
{ellipsoid_1945} simplifies to $N_{xx}\simeq{d}/({h+d})$ and $N_{yy}\simeq{h}%
/({h+d})$ \cite{Kittel_book,Kittel_1948,Ding_nanowire}, while%
\begin{equation}
\omega_{K}^{\left( W\right) }\rightarrow\omega_{K}^{\left( 0\right) }\left(
1+\frac{1}{2}\left( \frac{\mu_{0}\gamma M_{0}}{\omega_{K}^{\left( 0\right) }}%
\right) ^{2}\frac{h}{d}\right) ,  \label{omW}
\end{equation}
where $\omega_{K}^{\left( 0\right) }=\mu_{0}\gamma\sqrt{\left( H_{\mathrm{app%
}}+M_{0}\right) H_{\mathrm{app}}}$ is the FMR frequency of the extended film.

Under FMR, the Kittel modes of all wires excited by a homogeneous microwave
field that precess in phase. The magnetization $\mathbf{M}(\mathbf{r})$ is
periodic in the direction perpendicular to the nanowires, 
\begin{equation}
\mathbf{M}(\mathbf{r,}t)=\left\{ 
\begin{array}{c}
\mathbf{M}\left( t\right) ,~~y\in\lbrack na-\frac{d}{2},na+\frac{d}{2}%
],~x\in\lbrack0,h] \\ 
\hspace{-3.14cm}0,~~~~~~~\mathrm{otherwise}%
\end{array}
\right. ,
\end{equation}
where $n$ is an integer. The Fourier series of the transverse components of $%
\mathbf{M}(\mathbf{r})$ reads 
\begin{equation}
{M}_{\beta}^{\mathrm{K}}(\mathbf{r},t)={m}_{\beta}^{\mathrm{K}}e^{-i\omega
_{K}t}\Theta(h-x)\Theta(x)\sum_{m\geq0}^{\mathrm{even}}2f_{m}\cos
k_{y}^{(m)}y,  \label{Kittel_wave}
\end{equation}
in which $\beta=\left\{ x,y\right\} $, $\Theta(x)$ is the Heavyside step
function, $k_{y}^{(m)}=m{\pi}/{a}$ with $m$ a positive \textit{even} integer
and 
\begin{equation}
f_{m}=\left( 1-\frac{1}{2}\delta_{m0}\right) \frac{2}{\pi m}\sin\left( \frac{%
d}{2}k_{y}^{\left( m\right) }\right) .
\end{equation}
${M}_{\beta}^{\mathrm{K}}(\mathbf{r})$ is the lowest acoustic mode with
frequency $\omega_{K}$ for the nanowire array in the interval $\omega
_{K}^{\left( 0\right) }\leq\omega_{K}\leq\omega_{K}^{\left( W\right) }$ with 
$\omega_{K}^{\left( W\right) }-\omega_{K}^{\left( 0\right) }=O\left(
h/d\right) $ \cite{crystal1,crystal2,magnonics1}. The normalization
condition of the amplitudes read (for general modes labelled by $p$) \cite%
{Walker,magnetic_nanodots} 
\begin{equation}
\int d\mathbf{r}[M_{x}^{\left( p\right) }(\mathbf{r})\overline {%
M_{y}^{\left( p\right) }(\mathbf{r})}-\overline{M_{x}^{\left( p\right) }(%
\mathbf{r})}M_{y}^{\left( p\right) }(\mathbf{r})]=-i/2,
\label{normalization2}
\end{equation}
where $\overline{M}=M^{\ast}$. The acoustic mode in Eq.~(\ref{Kittel_wave})
is elliptically polarized as 
\begin{align}
{m}_{x}^{\mathrm{K}} & =\sqrt{\frac{a}{4hd}\sqrt{\frac{H_{\mathrm{app}
}+M_{0}N_{yy}}{H_{\mathrm{app}}+M_{0}N_{xx}}}}  \notag \\
&\rightarrow\frac{1}{2d}\sqrt{\frac{ad}{h}\sqrt{\frac{H_{\mathrm{app}}}{H_{%
\mathrm{app}}+M_{0}}}}+O\left( \frac{h}{d}\right),  \notag
\end{align}
\begin{align}
{m}_{y}^{\mathrm{K}} & =i\sqrt{\frac{a}{4hd}\sqrt{\frac{H_{\mathrm{app}
}+M_{0}N_{xx}}{H_{\mathrm{app}}+M_{0}N_{yy}}}}  \notag \\
&\rightarrow\frac{i}{2d}\sqrt{\frac{ad}{h}\sqrt{\frac{H_{\mathrm{app}}+M_{0}%
}{H_{\mathrm{app}}}}}+O\left( \frac{h}{d}\right),  \label{wave_nanowire}
\end{align}
which can be strongly elliptic in the thin film limit.

\subsection{Spin waves in a thin magnetic film}

\label{spin_waves_film}

Magnetic modes $\mathbf{\tilde{M}}$ in the film are the solution of the
Landau-Lifshitz (LL) equation \cite{Landau} 
\begin{equation}
{d}\mathbf{\tilde{M}}/{dt}=-\mu_{0}\gamma\mathbf{\tilde{M}}\times (\mathbf{H}%
_{\mathrm{app}}+\mathbf{\tilde{H}}_{d}+\mathbf{\tilde{H}}_{\mathrm{ex}}),
\label{EOM}
\end{equation}
where $\mathbf{H}_{\mathrm{app}}=H_{\mathrm{app}}\mathbf{\hat{z}}$ is the
same applied magnetic field as above, $\mathbf{\tilde{H}}_{d}$ is the
dipolar field (see Appendix~\ref{BB}) \cite{Kalinikos}, and the exchange
field $\mathbf{\tilde{H}}_{\mathrm{ex}}=\alpha_{\mathrm{ex}}\left(
\partial_{x}^{2}\mathbf{\tilde{M}}+\partial_{y}^{2}\mathbf{\tilde{M}}%
+\partial_{z}^{2}\mathbf{\tilde{M}}\right) $ with stiffness $\alpha_{\mathrm{%
ex}}$. We choose free boundary conditions $d\mathbf{\tilde{M}}(\mathbf{r}%
)/dx|_{x=0,-s}=0$ for simplicity \cite{inplane1,inplane2,inplane3}, since
the lowest mode in sufficiently thin films is not affected by partial
pinning \cite{exchange_1969,exchange_six_order,Stamps_0,Stamps}.

By translational symmetry in the $\hat{\mathbf{y}}$-$\hat{\mathbf{z}}$
plane, ${\tilde{M}}_{x,y}(\mathbf{r})={\tilde{m}}_{x,y}^{\mathbf{k}%
}(x)e^{ik_{y} y+ik_{z}z}e^{-i\omega t}$ with $\mathbf{k}\equiv k_{y}\hat{%
\mathbf{y}}+k_{z}\hat{\mathbf{z}}$. We focus on the spin waves with $k_{z}=0$
that couple to the acoustic mode of the nanowire array (see Sec.~\ref%
{interlayer_coupling}). From Eqs.~(\ref{EOM}) and (\ref{Green}), ${\tilde{m}}%
_{\pm}^{k_{y}}(x)={\tilde{m}}_{x}^{k_{y}}(x)\pm i{\tilde{m}}_{y}^{k_{y}}(x) $
and $d{\tilde{m}}_{\pm}^{k_{y}}(x)/dx|_{x=0,-s}=0,$ we have the Fourier
series \cite{Kalinikos,Kostylev} 
\begin{equation}
{\tilde{m}}_{\pm}^{k_{y}}(x)=\sum_{l=0}^{\infty}(\sqrt{2}/\sqrt{1+\delta_{l0}%
}){\tilde{m}}_{l,\pm}^{k_{y}}\cos\frac{l\pi x}{s}.
\end{equation}
Eq.~(\ref{EOM}) leads to the following equations for ${\tilde{m}}%
_{l,\pm}^{k_{y}}$ \cite{Kalinikos,Kostylev}, 
\begin{widetext}
\begin{gather}
\left( 
\begin{array}{cc}
\tilde{\omega}+\Omega_{H}+\alpha_{\mathrm{ex}}k_{y}^{2}+\alpha_{\mathrm{ex}%
}\left( {l\pi}/{s}\right) ^{2}+{1}/{2} & {1}/{2}-|k_{y}|Q_{ll}/2 \\ 
-{1}/{2}+|k_{y}|Q_{ll}/2 & \tilde{\omega}-\Omega_{H}-\alpha_{\mathrm{ex}%
}k_{y}^{2}-\alpha_{\mathrm{ex}}\left( {l\pi}/{s}\right) ^{2}-{1}/{2}%
\end{array}
\right) \left( 
\begin{array}{c}
\tilde{{m}}_{l,+}^{k_{y}} \\ 
\tilde{{m}}_{l,-}^{k_{y}}%
\end{array}
\right) +  \notag \\
+\sum_{l^{\prime}\neq l}\left( 
\begin{array}{cc}
0 & -k_{y}\tilde{Q}_{ll^{\prime}}/2-|k_{y}|Q_{ll^{\prime}}/2 \\ 
-k_{y}\tilde{Q}_{ll^{\prime}}/2+|k_{y}|Q_{ll^{\prime}}/2 & 0%
\end{array}
\right) \left( 
\begin{array}{c}
\tilde{{m}}_{l^{\prime},+}^{k_{y}} \\ 
\tilde{{m}}_{l^{\prime},-}^{k_{y}}%
\end{array}
\right) =0,  \label{eigen_full}
\end{gather}
where $\tilde{\omega}\equiv\omega/(\mu_{0}\gamma\tilde{M}_{0})$, $\Omega
_{H}\equiv H_{\mathrm{app}}/\tilde{M}_{0},$ and 
\begin{align}
&Q_{ll^{\prime}}=\frac{1}{s}\int_{-s}^{0}dx\int_{-s}^{0}dx^{\prime
}e^{-\left\vert x-x^{\prime}\right\vert \left\vert k_{y}\right\vert
}\cos\left( \frac{l^{\prime}\pi}{s}x^{\prime}\right) \cos\left( \frac{l\pi }{%
s}x\right)\frac{2}{\sqrt{(1+\delta_{l0})(1+\delta_{l^{\prime}0})}},  \notag
\\
&\tilde{Q}_{ll^{\prime}}=\frac{1}{s}\int_{-s}^{0}dx\int_{-s}^{0}dx^{\prime}%
\mathrm{sgn}\left( x-x^{\prime|}\right) e^{-\left\vert
x-x^{\prime}\right\vert \left\vert k_{y}\right\vert }\cos\left( \frac{%
l^{\prime}\pi}{s}x^{\prime}\right)\cos\left( \frac{l\pi}{s}x\right) \frac{2}{%
\sqrt{(1+\delta _{l0})(1+\delta_{l^{\prime}0})}},  \label{QQs}
\end{align}
and $\mathrm{sgn}(x-x^{\prime})=1$ when $x>x^{\prime}$ and $\mathrm{sgn}%
(x-x^{\prime})=-1$ when $x<x^{\prime}$. 
\end{widetext}

The exchange energy for the spin waves along the $\hat{\mathbf{x}}$%
-direction is $\alpha _{\mathrm{ex}}\left( {l\pi }/{s}\right) ^{2}$. For the
typical film thickness $s\leq 20$~nm and magnon wavelength $2\pi
/k_{y}\gtrsim 100$~nm, $\alpha _{\mathrm{ex}}k_{y}^{2}\ll \alpha _{\mathrm{ex%
}}\left( {l\pi }/{s}\right) ^{2}$ when $l\geq 1$. In Appendix~\ref{CC}, we
argue that we may confine our attention to the spin waves in the lowest
branch ${l=0}$ with amplitude governed by \cite{Kalinikos,Kostylev} 
\begin{widetext}
\begin{equation}
\omega _{0}\left( 
\begin{array}{c}
\tilde{{m}}_{0,+}^{k_{y}} \\ 
\tilde{{m}}_{0,-}^{k_{y}}%
\end{array}%
\right) =\mu _{0}\gamma \tilde{M}_{0}\left( 
\begin{array}{cc}
-\Omega _{H}-\alpha _{\mathrm{ex}}k_{y}^{2}-\frac{1}{2} & \frac{1}{2}-\frac{1%
}{s|k_{y}|}\left( 1-\frac{1}{s|k_{y}|}e^{-s|k_{y}|}\right) \\ 
-\frac{1}{2}+\frac{1}{s|k_{y}|}\left( 1-\frac{1}{s|k_{y}|}%
e^{-s|k_{y}|}\right) & \Omega _{H}+\alpha _{\mathrm{ex}}k_{y}^{2}+\frac{1}{2}%
\end{array}%
\right) \left( 
\begin{array}{c}
\tilde{{m}}_{0,+}^{k_{y}} \\ 
\tilde{{m}}_{0,-}^{k_{y}}%
\end{array}%
\right) ,  \label{homogeneous}
\end{equation}%
\end{widetext}
which leads to the energy spectrum \cite%
{Kalinikos,inplane1,inplane2,inplane3,Hurben} 
\begin{align}
\omega _{0}& =\mu _{0}\gamma \tilde{M}_{0}\left[ (\Omega _{H}+\alpha _{%
\mathrm{ex}}k_{y}^{2}+1)(\Omega _{H}+\alpha _{\mathrm{ex}}k_{y}^{2})\right. 
\notag \\
& \left. +\left( 1-\frac{1}{|k_{y}|s}+\frac{1}{|k_{y}|s}e^{-|k_{y}|s}\right)
\left( \frac{1}{|k_{y}|s}-\frac{1}{|k_{y}|s}e^{-|k_{y}|s}\right) \right] ^{%
\frac{1}{2}}.
\end{align}%
and ellipticity 
\begin{equation}
{\tilde{m}}_{0,y}^{k_{y}}=i\frac{F-1}{F+1}{\tilde{m}}_{0,x}^{k_{y}},
\label{phase_d}
\end{equation}%
where 
\begin{equation}
F=\frac{-\frac{{1}}{{2}}+\frac{1-\exp {(-|k_{y}|s)}}{{|k_{y}|s}}}{\frac{%
\omega _{0}}{\mu _{0}\gamma \tilde{M}_{0}}-(\Omega _{H}+\alpha _{\mathrm{ex}%
}k_{y}^{2}+1/2)}.
\end{equation}%
With the normalization Eq.~(\ref{normalization2}) we find 
\begin{equation}
{\tilde{m}}_{0,x}^{k_{y}}=\sqrt{\frac{{F+1}}{{4s\left( {F-1}\right) }}},~~~~{%
\tilde{m}}_{0,y}^{k_{y}}=i\sqrt{\frac{{F-1}}{{4s}\left( {F+1}\right) }}.
\label{waves_film}
\end{equation}%
For wavelengths that are relatively short but still much larger than the
film thickness or $s/\alpha _{\mathrm{ex}}\lesssim |k_{y}|\lesssim 1/s$, the
energy of the spin waves in the lowest branch approaches $\mu _{0}\gamma 
\tilde{M}_{0}\sqrt{(\Omega _{H}+\alpha _{\mathrm{ex}}k_{y}^{2}+1)(\Omega
_{H}+\alpha _{\mathrm{ex}}k_{y}^{2})}$, $\left\vert F\right\vert \gg 1$ and
the precession becomes circular with ${\tilde{m}}_{0,y}^{k_{y}}=i{\tilde{m}}%
_{0,x}^{k_{y}}=i\sqrt{1/(4s)}$.

\section{Interlayer dipolar and exchange interactions}

\label{interlayer_coupling} We now analyze the coupling between the NWA and
film that generate the observed anticrossings in the microwave absorption.
We focus on the experimental relevant interlayer dipolar and exchange
couplings of the Kittel mode of the NWA and the spin waves in the lowest
subband of the thin film. We adopt the configuration in which the
equilibrium magnetizations and applied field are all parallel to the $\hat{%
\mathbf{z}}$-direction. The results also hold for the antiparallel
configuration with $\mathbf{\tilde{m}}_{0}\parallel \mathbf{H}_{\mathrm{app}%
}\parallel \hat{\mathbf{z}}$ and $\mathbf{M}_{0}\parallel -\hat{\mathbf{z}}$%
, by replacing $m_{y}^{\mathrm{K}}$ with $-m_{y}^{\mathrm{K}}$.

\subsection{Interlayer dipolar interaction}

\label{interlayer_dipolar_interaction}

The free energy due to the interlayer dipolar interaction reads \cite{Landau}
\begin{eqnarray}
F_{d}(t)&=&-\mu_{0}\int\mathbf{\tilde{M}}(\mathbf{r},t)\cdot\mathbf{h}^{D}(%
\mathbf{r},t)d\mathbf{r}  \notag \\
&=&-\mu_{0}\int\mathbf{M}^{\mathrm{K}}(\mathbf{r},t)\cdot\mathbf{\tilde{h}}%
^{D}(\mathbf{r},t)d\mathbf{r},  \label{dipolar_energy}
\end{eqnarray}
where $\mathbf{h}^{D}$ ($\mathbf{\tilde{h}}^{D}$) is the demagnetization
field generated by the acoustic mode (spin waves) in the NWA (films) \cite%
{Landau} 
\begin{eqnarray}
&&{h}_{\beta}^{D}(\mathbf{r},t)=\frac{1}{4\pi}\partial_{\beta}\int d\mathbf{r%
}^{\prime}\frac{\partial_{\alpha}{M}_{\alpha}^{\mathrm{K} }\mathbf{%
(r^{\prime},t)}}{|\mathbf{r}-\mathbf{r}^{\prime}|},  \notag \\
&&{\tilde{h}}_{\beta}^{D}(\mathbf{r},t)=\frac{1}{4\pi}\partial_{\beta}\int d%
\mathbf{r}^{\prime}\frac{\partial_{\alpha}{\tilde{M}}_{\alpha}\mathbf{%
(r^{\prime},t)}}{|\mathbf{r}-\mathbf{r}^{\prime}|},  \label{dipolar_fields}
\end{eqnarray}
where $\alpha,\beta=\left\{ x,y\right\} $ and the repeated index implies
summation (over $\alpha$). The field acting on the nanowire array is $%
\mathbf{\tilde{h}}^{D}$ with $x>0>x^{\prime}$ (see Fig.~\ref{grating1})
while that acting on the film is $\mathbf{h}^{D}$ with $x<0$. Below, rich
features are revealed for the interlayer dipolar coupling by both classical
and quantum descriptions, which are further understood from the unique
behaviors of $\mathbf{h}^{D}$ and $\mathbf{\tilde{h}}^{D}$.

\subsubsection{Classical description}

In a classical description, the Kittel mode of the NWA as derived above
reads (see Eq.~(\ref{Kittel_wave})) 
\begin{eqnarray}
&&\left( 
\begin{array}{c}
{M}_{x}^{\mathrm{K}}(\mathbf{r},t) \\ 
{M}_{y}^{\mathrm{K}}(\mathbf{r},t)%
\end{array}%
\right) =\Theta (h-x)\Theta (x)  \notag \\
&&\times \sum_{m\geq 0}^{\mathrm{even}}2f_{m}\cos (k_{y}^{(m)}y)\left( 
\begin{array}{c}
{m}_{x}^{\mathrm{K}}\cos (\omega _{K}t) \\ 
{m}_{y}^{\mathrm{K}}\sin (\omega _{K}t)%
\end{array}%
\right) .
\end{eqnarray}%
By substituting these modes into Eq.~(\ref{dipolar_fields}) and using the
Coulomb integral 
\begin{equation}
I=\int d\mathbf{r}^{\prime }\frac{e^{ik_{y}y^{\prime }}f\left( x^{\prime
}\right) }{|\mathbf{r}-\mathbf{r}^{\prime }|}=\frac{2\pi }{|k_{y}|}%
e^{ik_{y}y}\int dx^{\prime }e^{-\left\vert x-x^{\prime }\right\vert
|k_{y}|}f\left( x^{\prime }\right) ,
\end{equation}%
its dipolar field in the film below becomes 
\begin{eqnarray}
&&\left( 
\begin{array}{c}
h_{x}^{D}(\mathbf{r},t) \\ 
h_{y}^{D}(\mathbf{r},t)%
\end{array}%
\right) =\sum_{m\geq 0}^{\mathrm{even}}F_{m}e^{\left\vert
k_{y}^{(m)}\right\vert x}  \notag \\
&&\mbox{}\times\left( 
\begin{array}{cc}
\cos \left( k_{y}^{(m)}y\right) & -\sin \left( k_{y}^{(m)}y\right) \\ 
-\sin \left( k_{y}^{(m)}y\right) & -\cos \left( k_{y}^{(m)}y\right)%
\end{array}%
\right) \left( 
\begin{array}{c}
{m}_{x}^{\mathrm{K}}\cos (\omega _{K}t) \\ 
{m}_{y}^{\mathrm{K}}\sin (\omega _{K}t)%
\end{array}%
\right) ,\notag\\
  \label{d_NWA}
\end{eqnarray}%
with the form factor $F_{m}=f_{m}(1-e^{-\left\vert k_{y}^{(m)}\right\vert
h}) $. By inspection of the dipolar fields under the wire center and between
the wires 
\begin{align}
&\mathbf{h}^{D}\left( x,y=0,t\right) =\sum_{m\geq 0}^{\mathrm{even}%
}F_{m}e^{|k_{y}^{(m)}|x}\left( 
\begin{array}{c}
m_{x}^{\mathrm{K}}\cos (-\omega _{K}t) \\ 
m_{y}^{\mathrm{K}}\sin (-\omega _{K}t)%
\end{array}%
\right) , \\
&\mathbf{h}^{D}\left( x,y=\frac{a}{2},t\right) =\sum_{m\geq 0}^{\mathrm{even%
}}F_{m}\left( -1\right) ^{\frac{m}{2}}e^{|k_{y}^{(m)}|x}\left( 
\begin{array}{c}
m_{x}^{\mathrm{K}}\cos (-\omega _{K}t) \\ 
m_{y}^{\mathrm{K}}\sin (-\omega _{K}t)%
\end{array}%
\right) ,
\end{align}%
it becomes clear that $\mathbf{h}^{D}$ rotates in the $x$-$y$ plane, but in
opposite direction of $\mathbf{M}^{\mathrm{K}}$. Decomposing the latter into
right and left circularly-polarized components as $%
(m_{x}^{K},m_{y}^{K})^{T}=m_{R}^{K}(1,1)^{T}+m_{L}^{K}(1,-1)^{T}$, the
dipolar field Eq.~(\ref{d_NWA}) can be written 
\begin{eqnarray}
\left( 
\begin{array}{c}
h_{x}^{D}(\mathbf{r}) \\ 
h_{y}^{D}(\mathbf{r})%
\end{array}%
\right) &=&\sum_{m\geq 0}^{\mathrm{even}}F_{m}e^{\left\vert
k_{y}^{(m)}\right\vert x}\left[ m_{R}^{K}\left( 
\begin{array}{c}
\cos (-k_{y}^{(m)}y-\omega _{K}t) \\ 
\sin (-k_{y}^{(m)}y-\omega _{K}t)%
\end{array}%
\right) \right.  \notag \\
&&\left. +m_{L}^{K}\left( 
\begin{array}{c}
\cos (k_{y}^{(m)}y-\omega _{K}t) \\ 
-\sin (k_{y}^{(m)}y-\omega _{K}t)%
\end{array}%
\right) \right] .  \label{chiral_dipolar}
\end{eqnarray}%
Since $k_{y}^{(m)}\geq 0$, the standing magnetization mode in the NWA
generates two travelling dipolar field waves with opposite direction locked
by the polarization. A right circularly polarized Kittel mode $\left(
m_{L}^{K}=0\right) $ generates dipolar magnetic fields with the opposite
polarization that propagate only in one direction, while ellipticity leads
to a second wave with same polarization sense but in opposite direction.

$\mathbf{h}^{D}(\mathbf{r})$ can now interact with the proximate spin waves
in the film below, which we denote as%
\begin{equation}
\mathbf{\tilde{M}}(\mathbf{r},t)=\left( 
\begin{array}{c}
\tilde{m}_{x}^{\mathbf{k}}(x)\cos \left( \mathbf{k}\cdot \mathbf{r}%
_{\parallel }-\omega t\right) \\ 
-\tilde{m}_{y}^{\mathbf{k}}(x)\sin \left( \mathbf{k}\cdot \mathbf{r}%
_{\parallel }-\omega t\right)%
\end{array}%
\right) ,  \label{Mtilde}
\end{equation}%
where $\mathbf{r}_{\parallel }=y\hat{\mathbf{y}}+z\hat{\mathbf{z}}$.
Substituting, the magnetic free energy due to the interlayer dipolar
coupling becomes 
\begin{align}
&F_{d}(t) =-\mu _{0}\sum_{m\geq 0}^{\mathrm{even}}F_{m}\int d\mathbf{r}%
e^{|k_{y}^{(m)}|x}  \notag \\
& \times \left( \tilde{m}_{x}^{\mathbf{k}}(x)\cos \left( \mathbf{k}\cdot 
\mathbf{r}_{\parallel }-\omega t\right) ,-\tilde{m}_{y}^{\mathbf{k}}(x)\sin
\left( \mathbf{k}\cdot \mathbf{r}_{\parallel }-\omega t\right) \right) 
\notag \\
& \times \left( 
\begin{array}{cc}
\cos \left( k_{y}^{(m)}y\right) & -\sin \left( k_{y}^{(m)}y\right) \\ 
-\sin \left( k_{y}^{(m)}y\right) & -\cos \left( k_{y}^{(m)}y\right)%
\end{array}%
\right) \left( 
\begin{array}{c}
m_{x}^{\mathrm{K}}\cos (\omega _{K}t) \\ 
m_{y}^{\mathrm{K}}\sin (\omega _{K}t)%
\end{array}%
\right) .  \label{free_dd}
\end{align}%
The dipolar thin-film form anisotropy also causes elliptical precessions
that can be decomposed into the right and left circularly-polarized
components as $\left( \tilde{m}_{x}^{\mathbf{k}}(x),\tilde{m}_{y}^{\mathbf{k}%
}(x)\right) =\tilde{m}_{R}^{\mathbf{k}}(x)(1,1)+\tilde{m}_{L}^{\mathbf{k}%
}(x)(1,-1)$. At resonance $\omega =\omega _{K}$ the average $\bar{F}_{d}$
over a time period $2\pi /\omega _{K}$ is finite%
\begin{eqnarray}
\bar{F}_{d} &=&-\mu _{0}\sum_{m\geq 0}^{\mathrm{even}}F_{m}\int
dxe^{|k_{y}^{(m)}|x}  \notag \\
&&\times \left( \tilde{m}_{R}^{k_{y}}(x)m_{L}^{K}\delta _{k_{y},k_{y}^{(m)}}+%
\tilde{m}_{L}^{k_{y}}(x)m_{R}^{K}\delta _{k_{y},-k_{y}^{(m)}}\right) .
\label{compact}
\end{eqnarray}%
Eq.~(\ref{compact}) leads to the following conclusions:

\begin{itemize}
\item The ac dipolar magnetic fields couple only to spin waves with the same
polarization (conservation of angular momentum).

\item The FMR resonance of the NWA couples only to spin waves with momentum $%
\pm k_{y}^{(m)}\hat{\mathbf{y}}$ (conservation of linear momentum).

\item Circularly polarized excitations in both NWA and film to do not
interact when equilibrium magnetizations are parallel. However, they do
couple in the antiparallel configuration, which is obtained from Eq.~(\ref%
{compact}) by exchanging $\tilde{m}_{R}^{k_{y}}\leftrightarrow\tilde{m}%
_{L}^{k_{y}}$.

\item When the spin waves are circularly polarized, i.e. $\tilde{m}_{L}
^{k_{y}}=0,$ but the NWA modes are elliptic, the coupling is perfectly
chiral, i.e. the Kittel mode of the NWA interacts with spin waves that
propagate in one direction only.

\item A finite chirality persists when both the spin waves and NWA mode are
elliptically polarized as long as $\tilde{m}_{R}^{k_{y}^{(m)}}(x)m_{L}^{K}
\neq\tilde{m}_{L}^{-k_{y}^{(m)}}(x)m_{R}^{K}$.
\end{itemize}

The dipolar (magnetostatic) spin waves in thin films are elliptically
polarized due to the anisotropy of demagnetization fields, with the
exception of the DE modes in thick films as discussed briefly below. The NWA
Kittel mode then asymmetrically mixes with spin waves in both directions. At
higher frequencies the dipolar interaction becomes less dominant and the
spin waves become nearly circularly polarized, $\tilde{m}_{L}^{k_{y}}%
\rightarrow0$, which implies that only spin waves propagating in one
direction interact as long as $m_{L}^{K}\neq0$. When the magnetizations are
antiparallel, $m_{R}^{K}$ and $m_{L}^{K}$ are exchanged, leading to perfect
and large chiral coupling for circularly polarized magnetization dynamics.

The physics can be also understood in terms of the dipolar field generated
by the spin waves and acting on the NWA. We can express the spin waves in
the thin film as 
\begin{eqnarray}
\left( 
\begin{array}{c}
\tilde{M}_{x}(\mathbf{r}) \\ 
\tilde{M}_{y}(\mathbf{r})%
\end{array}%
\right) &=&\tilde{m}_{R}(x)\left( 
\begin{array}{c}
\cos (k_{y}y-\omega t) \\ 
-\sin (k_{y}y-\omega t)%
\end{array}%
\right)  \notag \\
&&+\tilde{m}_{L}(x)\left( 
\begin{array}{c}
\cos (k_{y}y-\omega t) \\ 
\sin (k_{y}y-\omega t)%
\end{array}%
\right) ,
\end{eqnarray}%
where $\tilde{m}_{R}(x)$ and $\tilde{m}_{L}(x)$ denote the right and left
circularly-polarized components. Above the film with $x>0>x^{\prime }$, 
\begin{eqnarray}
&&\left( 
\begin{array}{c}
\tilde{h}_{x}^{D}(\mathbf{r}) \\ 
\tilde{h}_{y}^{D}(\mathbf{r})%
\end{array}%
\right)=\frac{1}{2}e^{-|k_{y}|x}\int dx^{\prime }\left[
(k_{y}+|k_{y}|)m_{R}(x^{\prime })\right.  \notag \\
&&\left. +(k_{y}-|k_{y}|)m_{L}(x^{\prime })\right] \left( 
\begin{array}{c}
\cos (k_{y}y-\omega t) \\ 
\sin (k_{y}y-\omega t)%
\end{array}%
\right) . 
 \label{circular_field}
\end{eqnarray}%
Irrespective of an ellipticity $\tilde{m}_{L}$, the dipolar field is left
(right) circularly polarized above (below) the film. Moreover, the dipolar
field generated by the right (left) circularly polarized components of the
spin waves does not vanish above the film only when $k_{y}>0$ ($k_{y}<0$).
This can be understood in terms of the surface magnetic charges with dipolar
fields that point in opposite direction on both sides of the film: When the
spin waves are right or left circularly polarized, only those travelling in
particular direction can couple with the NWA\ Kittel mode with fixed
circularly polarized component.

Although not treated here explicitly, we can draw some conclusions about the
DE modes in thick films as well. DE modes propagating perpendicular to the
magnetization can be excited efficiently by interlayer dipolar coupling
because they are circularly polarized, but the excitation efficiency is very
different for the parallel and anti-parallel configurations. Here, we
disregard the DE modes on the opposite side completely now since the film is
thick. From Eq.~(\ref{chiral_dipolar}), the anisotropic NWA generates the
right (left) circularly polarized magnetic fields propagating in (opposite
to) the $\hat{\mathbf{y}}$-direction determined by $m_{L}^{K}$ ($m_{R}^{K}$%
). With this in mind, DE modes of thick films can be efficiently excited by
dipolar interactions because they are confined to a thin skin near the
surface. However, the exchange coupling can also do that, irrespective of
the parallel vs. antiparallel configuration but with equal excitation
efficiency. The NWA therefore can be an efficient coupler to excite
short-wavelength DE modes that by the exchange interaction acquire a
significant group velocity.

\subsubsection{Quantum description}

We now formulate the interlayer dipolar coupling in second quantization
deriving the appropriate matrix elements from the classical interactions. In
order to make better contact with the literature, we replace the
magnetization $\mathbf{M}(\mathbf{r})$ by the spin operators $\hat{\mathbf{S}%
}(\mathbf{r})$ via $\mathbf{M}(\mathbf{r})\rightarrow -\gamma \hbar \hat{%
\mathbf{S}}(\mathbf{r)}$. After performing the Holstein-Primakoff
transformation \cite{Kittel_book,HP}, we linearize the problem in the magnon
operators and diagonalize the resulting Hamiltonian by a Bogoliubov
transformations \cite{squeezed_magnon,Kittel_book,HP,Sanchar_PRB}. The
leading term of the interaction between NWA and film then reads 
\begin{equation}
\hat{H}_{\mathrm{d}}=-\mu _{0}\gamma ^{2}\hbar ^{2}\frac{1}{4\pi }\int d%
\mathbf{r}\hat{\tilde{\mathbf{S}}}_{\alpha }(\mathbf{r})\partial _{\beta
}\int d\mathbf{r}^{\prime }\frac{\partial _{\alpha }{\hat{\mathbf{S}}%
_{\alpha }^{\mathrm{K}}(\mathbf{r^{\prime }})}}{|\mathbf{r}-\mathbf{r}%
^{\prime }|}.  \label{dipolar_1}
\end{equation}%
with spin operators (for the time being for general film thickness) \cite%
{squeezed_magnon,Kittel_book,HP}, 
\begin{eqnarray}
\hat{\tilde{S}}_{\gamma }(\mathbf{r},t) &=&\sqrt{2\tilde{S}}\sum_{j\mathbf{k}%
}\left( \tilde{M}_{\gamma }^{\left( j\mathbf{k}\right) }(\mathbf{r})\hat{%
\alpha}_{j\mathbf{k}}(t)+\overline{\tilde{M}_{\gamma }^{\left( j\mathbf{k}%
\right) }}(\mathbf{r})\hat{\alpha}_{j\mathbf{k}}^{\dagger }(t)\right) , 
\notag \\
\hat{S}_{\delta }(\mathbf{r},t) &=&\sqrt{2S}\sum_{p}\left( M_{\delta
}^{\left( p\right) }(\mathbf{r})\hat{\beta}_{p}(t)+\overline{M_{\delta
}^{\left( p\right) }}(\mathbf{r})\hat{\beta}_{p}^{\dagger }(t)\right) ,
\label{expansion}
\end{eqnarray}%
with $\gamma ,\delta =\left\{ x,y\right\} $. Here $\hat{\alpha}_{j\mathbf{k}%
} $ is a magnon annihilation operator with band index $j$ in the film and
the Kittel mode of the nanowire array is annihilated by $\hat{\beta}_{K}$.
Then 
\begin{align}
\hat{H}_{\mathrm{d}}& =-\mu _{0}\gamma \hbar ^{2}\sqrt{\tilde{M}_{0}M_{0}} 
\notag \\
& \times \sum_{j\mathbf{k}}\left( \tilde{B}_{j\mathbf{k},\mathrm{K}}\hat{%
\beta}_{j\mathbf{k}}\hat{\alpha}_{\mathrm{K}}^{\dagger }+\tilde{A}_{j\mathbf{%
k},\mathrm{K}}\hat{\beta}_{j\mathbf{k}}\hat{\alpha}_{\mathrm{K}}+\mathrm{h.c.%
}\right) ,
\end{align}%
in terms of 
\begin{align}
\tilde{B}_{j\mathbf{k},\mathrm{K}}& =\sum_{m}F_{m}\int d\mathbf{r}%
e^{k_{y}^{(m)}x}\mathcal{\tilde{M}}_{j\mathbf{k}}(\mathbf{r})\left( \mathcal{%
Q}_{k_{y}^{(m)}}+\overline{\mathcal{Q}_{k_{y}^{(m)}}}\right) \overline{%
\mathcal{N}^{K}},  \notag \\
\tilde{A}_{j\mathbf{k},\mathrm{K}}& =\sum_{m}F_{m}\int d\mathbf{r}%
e^{k_{y}^{(m)}x}\mathcal{\tilde{M}}_{j\mathbf{k}}(\mathbf{r})(\mathcal{Q}%
_{k_{y}^{(m)}}+\overline{\mathcal{Q}_{k_{y}^{(m)}}})\mathcal{N}^{K}.
\label{coupling}
\end{align}%
Here, $F_{m}=f_{m}(1-e^{-k_{y}^{(m)}h})$, ${\mathcal{\tilde{M}}}_{j\mathbf{k}%
}(\mathbf{r})=\left( \tilde{M}_{x}^{\left( j\mathbf{k}\right) }(\mathbf{r}),%
\tilde{M}_{y}^{\left( j\mathbf{k}\right) }(\mathbf{r})\right) $, ${\mathcal{N%
}}^{K}=(m_{x}^{\mathrm{K}},m_{y}^{\mathrm{K}})^{T}$, and 
\begin{equation}
\mathcal{Q}_{k_{y}^{(m)}}=e^{ik_{y}^{(m)}y}\left( 
\begin{array}{cc}
1 & i \\ 
i & -1%
\end{array}%
\right) .
\end{equation}%
and calligraphic letters denote matrices here and below.

Only terms with $\left\vert \mathbf{k}_{y}\right\vert =k_{y}^{(m)}$ survive
the spatial integration in Eq.~(\ref{coupling}), which reflects momentum
conservation. In the following, we focus again on the experimental relevant
regime \cite{Haiming_NC, Haiming_PRL} of spin waves in the lowest branch $%
j=0,$ labeled in the following by \textquotedblleft H\textquotedblright\ and
the acoustic mode \textquotedblleft K\textquotedblright\ in the nanowire
array. With $\tilde{M}_{\beta}^{\mathrm{H,}\mathbf{k}}=\tilde{m}_{0,{\beta}%
}^{\mathbf{k}}e^{i\mathbf{k}\cdot\mathbf{r}}$ 
\begin{align}
\hat{H}_{\mathrm{d}} & =\sum_{m}\left( D_{d}^{m}\hat{\beta}_{\mathrm{H}
,-k_{y}^{(m)}}\hat{\alpha}_{\mathrm{K}}^{\dagger}+C_{d}^{m}\hat{\beta }_{%
\mathrm{H},k_{y}^{(m)}}\hat{\alpha}_{\mathrm{K}}^{\dagger}\right.  \notag \\
& \left. +A_{d}^{m}\hat{\beta}_{\mathrm{H},-k_{y}^{(m)}}\hat{\alpha }_{%
\mathrm{K}}+B_{d}^{m}\hat{\beta}_{\mathrm{H},k_{y}^{(m)}}\hat{\alpha }_{%
\mathrm{K}}+\mathrm{h.c.}\right) ,
\end{align}
in which 
\begin{align}
D_{d}^{m} & =-\mu_{0}\gamma\hbar^{2}\sqrt{\tilde{M}_{0}M_{0}}F_{m}\int
dxe^{k_{y}^{(m)}x}{\mathcal{P}}_{-k_{y}^{(m)}}(x)\mathcal{T}\overline {%
\mathcal{N}^{K}},  \notag \\
C_{d}^{m} & =-\mu_{0}\gamma\hbar^{2}\sqrt{\tilde{M}_{0}M_{0}}F_{m}\int
dxe^{k_{y}^{(m)}x}{\mathcal{P}}_{k_{y}^{(m)}}(x)\overline{\mathcal{T}}%
\overline{\mathcal{N}^{K}},  \notag \\
A_{d}^{m} & =-\mu_{0}\gamma\hbar^{2}\sqrt{\tilde{M}_{0}M_{0}}F_{m}\int
dxe^{k_{y}^{(m)}x}{\mathcal{P}}_{-k_{y}^{(m)}}(x)\mathcal{T}\mathcal{N}^{K},
\notag \\
B_{d}^{m} & =-\mu_{0}\gamma\hbar^{2}\sqrt{\tilde{M}_{0}M_{0}}F_{m}\int
dxe^{k_{y}^{(m)}x}{\mathcal{P}}_{k_{y}^{(m)}}(x)\overline{\mathcal{T}}%
\mathcal{N}^{K},  \label{dipolar_couplings}
\end{align}
with the spinor 
\begin{equation}
{\mathcal{P}}_{\pm k_{y}^{(m)}}(x)=\left( \tilde{m}_{0,x}^{\pm
k_{y}^{(m)}}(x),\tilde{m}_{0,y}^{\pm k_{y}^{(m)}}(x)\right)
\end{equation}
and 
\begin{equation}
\mathcal{T}=\left( 
\begin{array}{cc}
1 & i \\ 
i & -1%
\end{array}
\right) .
\end{equation}
The equilibrium magnetization of the NWA and film are parallel to the $\hat{%
\mathbf{z}}$-direction. When they are antiparallel with $\mathbf{\tilde 	{m}}%
_{0}\parallel\mathbf{H}_{\mathrm{app}}\parallel\hat{\mathbf{z}}$ and $%
\mathbf{M}_{0}\parallel-\hat{\mathbf{z}}$, $m_{y}^{\mathrm{K}}$ should be
replaced by $-m_{y}^{\mathrm{K}}$, as before.

We emphasize again that the couplings of the spin waves of opposite momentum
to the acoustic Kittel mode in the NWA can be very different. When the
wavelength is relatively short, or $\left\vert k_{y}\right\vert \gtrsim
s/\alpha_{\mathrm{ex}},$ the spin waves are nearly circularly polarized, $%
\tilde{m}_{0,y}^{\pm k_{y}^{(m)}}\approx i\tilde{m}_{0,x}^{\pm k_{y}^{(m)}}$%
. When substituted into $D_{d}^{m}$, the integral%
\begin{equation}
\int dxe^{k_{y}^{(m)}x}\tilde{m}_{0,x}^{-k_{y}^{(m)}}(1,i)\left( 
\begin{array}{cc}
1 & i \\ 
i & -1%
\end{array}
\right) \left( 
\begin{array}{c}
\overline{m}_{x}^{K} \\ 
\overline{m}_{y}^{K}%
\end{array}
\right) =0
\end{equation}
and $D_{d}^{m}\approx A_{d}^{m}\approx0$ in Eq.~(\ref{dipolar_couplings}).
This implies that the dipolar interaction cannot couple spin waves with
momentum $-\left\vert k_{y}^{(m)}\right\vert \hat{\mathbf{y}}$ to the
acoustic mode in the nanowire, while such a restriction does not hold for
waves with $+\left\vert k_{y}^{(m)}\right\vert \hat{\mathbf{y}}$. In other
words, the microwave field couples to short-wavelength spin waves in thin
films via a nanowire grating in a chiral manner.

As discussed above, the physical reason for this unexpected selection rule
is the asymmetry of the dipolar field generated by (circularly polarized)
spin waves propagating normal to the magnetization ($\Vert\hat{\mathbf{y}}$%
). For a particular momentum $q_{y}\hat{\mathbf{y}}$, the dipolar field
generated by the circular spin waves on the upper side is 
\begin{eqnarray}
&&\left( 
\begin{array}{c}
\tilde{h}_{x}^{D}(\mathbf{r}) \\ 
\tilde{h}_{y}^{D}(\mathbf{r})%
\end{array}
\right) =\frac{e^{-|\mathbf{q}|x}}{2}e^{-i\omega_{\mathbf{q}}t}\int
dx^{\prime}e^{-\left\vert \mathbf{q}\right\vert x^{\prime}}\left( 
\begin{array}{cc}
|q_{y}| & -iq_{y} \\ 
-iq_{y} & -|q_{y}|%
\end{array}
\right)  \notag \\
&&\mbox{}\times \left( 
\begin{array}{c}
1 \\ 
i%
\end{array}
\right) \tilde{M}_{x}(x^{\prime},y,z),  \label{dipolar_2}
\end{eqnarray}
which vanishes for negative $q_{y}$ but is finite for positive $q_{y}$.
Therefore only spin waves with positive (negative) $q_{y}$ can couple (not
couple) with the magnetization in the nanowire array.

\begin{figure}[th]
\begin{center}
{\includegraphics[width=8.8cm]{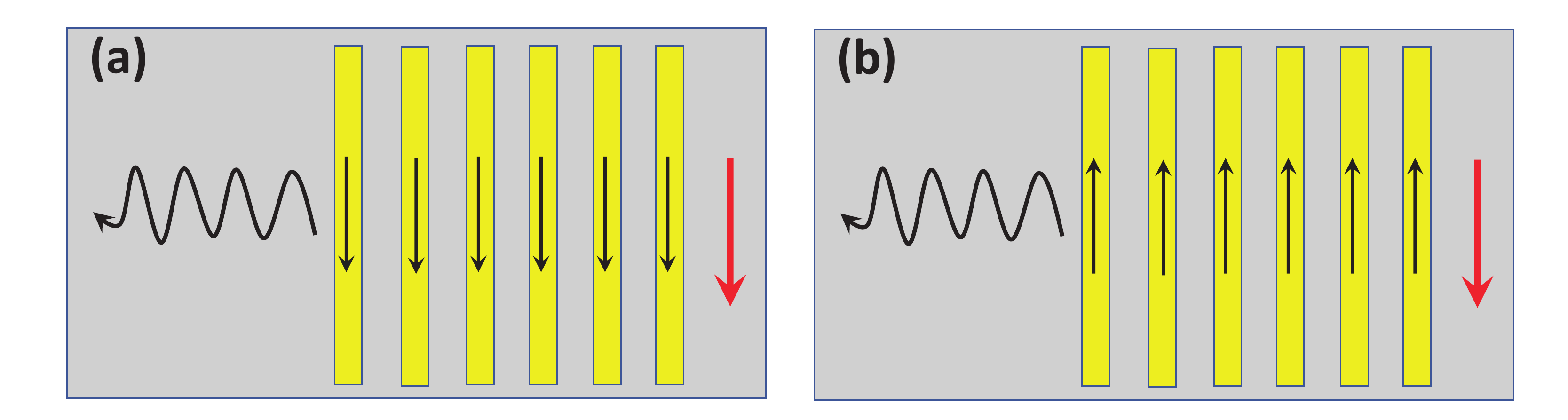}}
\end{center}
\caption{Chiral coupling of spin waves due to the interlayer dipolar
interaction for parallel and antiparallel magnetizations. The gray and
yellow regions denote the film and nanowire array. The red and black arrows
represent the direction of the soft magnetizations of the film in parallel to the
external field and NWA, respectively. The wavy line with arrow indicates the
propagating direction of spin waves that couple to the Kittel mode of the
NWA.}
\label{P_AP}
\end{figure}

The different excitation configurations and the chiral coupling are
illustrated by Fig.~\ref{P_AP}. When the nanowire array is fabricated on the
upper surface of the film, irrespective of whether the magnetizations in the
film and nanowires are parallel [Fig.~\ref{P_AP}(a)] or anti-parallel [Fig.~%
\ref{P_AP}(b)], among the short-wavelength spin waves only those with
momentum $\mathbf{k}\parallel(\mathbf{\tilde{m}}_{0}\times\mathbf{n})$
(shown by the wavy line with arrow) couple to (are excited by) the acoustic
NWA mode.

\subsection{NWA-magnetic film exchange interaction}

\label{interlayer_exchange_interaction}

Both the static \cite{static_ex1,static_ex2,static_t} and dynamic \cite%
{spin_pumping,Klingler12,Co_YIG,CoFeB_YIG,spin_diffusion} interlayer
exchange interaction between the NWA and magnetic film can play a role in
the coupling of short-wavelength spin waves in magnetic bilayers \cite%
{Klingler12,Co_YIG,CoFeB_YIG}. Here we focus on the dynamic exchange
interaction, eventually moderated by spin diffusion in a spacer layer \cite%
{spin_pumping,Heinrich,Co_YIG,CoFeB_YIG,spin_diffusion}. Indeed, recent
experiments \cite{Co_YIG} show that for direct contact between Co and YIG
bilayers, the static interfacial exchange interaction plays a dominant role
by locking the interface magnetization on both sides together. A 5-nm Cu
space layer, on the other hand, completely suppresses the static exchange
interaction, while the dynamic interaction mediated by the exchange of
non-equilibrium spin currents through the spacer remains \cite{Co_YIG}. A
1.5-nm AlO$_{x}$ layer suppresses both static and dynamic exchange
interactions \cite{Co_YIG}. Here, we assess the role of a significant direct
exchange interaction between the NWA and film, but do not discuss the
dissipative dynamic coupling.

The free energy due to an interfacial exchange interaction density $J$ can
be written as \cite{static_ex1,static_ex2,static_t,Co_YIG} 
\begin{align}
F_{\mathrm{ex}}& =\int d\mathbf{r}J\delta (x)\mathbf{\tilde{M}}(\mathbf{r}%
)\cdot \mathbf{M}(\mathbf{r})  \notag \\
& =J\int dydz\mathbf{\tilde{M}}(x=0,y,z)\cdot \mathbf{M}(x=0,y,z),
\label{ex_static}
\end{align}%
When $J>0$ ($J<0$), the interlayer exchange interaction is
anti-ferromagnetic (ferromagnetic). $J$ can be calculated by first
principles \cite{Jia_first_principle} or fitted to experiments \cite%
{static_ex1,static_ex2,Co_YIG,CoFeB_YIG,Haiming_PRL}. 
\begin{equation}
\hat{H}_{\mathrm{ex}}=J\gamma ^{2}\int dydz\hat{{\tilde{\mathbf{S}}}}%
(x=0,y,z)\cdot \hat{\mathbf{S}}(x=0,y,z).
\end{equation}%
As above, $\hat{{\tilde{\mathbf{S}}}}(x=0,y,z)$ and $\hat{\mathbf{S}}%
(x=0,y,z)$ represent the lowest spin wave subband in the film and the Kittel
mode of the NWA. The expansion into normal modes, Eq.~(\ref{expansion}), 
\begin{align}
\hat{H}_{\mathrm{ex}}& =\sum_{m}\left\{ D_{\mathrm{ex}}^{m}\hat{\beta}_{%
\mathrm{H}-k_{y}^{(m)}}\hat{\alpha}_{\mathrm{K}}^{\dagger }+C_{\mathrm{ex}%
}^{m}\hat{\beta}_{\mathrm{H}k_{y}^{(m)}}\hat{\alpha}_{\mathrm{K}}^{\dagger
}\right.  \notag \\
& \left. +A_{\mathrm{ex}}^{m}\hat{\beta}_{\mathrm{H}-k_{y}^{(m)}}\hat{\alpha}%
_{\mathrm{K}}+B_{\mathrm{ex}}^{m}\hat{\beta}_{\mathrm{H}-k_{y}^{(m)}}\hat{%
\alpha}_{\mathrm{K}}+\mathrm{h.c.}\right\} ,
\end{align}%
then contains the coefficients 
\begin{align}
D_{\mathrm{ex}}^{m}& =2J\sqrt{\tilde{M}_{0}M_{0}}f_{m}\sum_{\beta =x,y}%
\tilde{m}_{\beta }^{-k_{y}^{(m)}}(x=0)\overline{m_{0,\beta }^{\mathrm{K}}},
\label{exchange_couplings1} \\
C_{\mathrm{ex}}^{m}& =2J\sqrt{\tilde{M}_{0}M_{0}}f_{m}\sum_{\beta =x,y}%
\tilde{m}_{\beta }^{k_{y}^{(m)}}(x=0)\overline{m_{0,\beta }^{\mathrm{K}}},
\label{exchange_couplings2} \\
A_{\mathrm{ex}}^{m}& =2J\sqrt{\tilde{M}_{0}M_{0}}f_{m}\sum_{\beta =x,y}%
\tilde{m}_{\beta }^{-k_{y}^{(m)}}(x=0)m_{0,\beta }^{\mathrm{K}},
\label{exchange_couplings3} \\
B_{\mathrm{ex}}^{m}& =2J\sqrt{\tilde{M}_{0}M_{0}}f_{m}\sum_{\beta =x,y}%
\tilde{m}_{\beta }^{k_{y}^{(m)}}(x=0)m_{0,\beta }^{\mathrm{K}}.
\label{exchange_couplings4}
\end{align}%
For short-wavelength spin waves with nearly constant amplitude across a thin
film, $D_{\mathrm{ex}}^{m}\approx C_{\mathrm{ex}}^{m}$ and $A_{\mathrm{ex}%
}^{m}\approx B_{\mathrm{ex}}^{m}$. The expressions above hold when
magnetizations in the NWA and films are both parallel to the $\hat{\mathbf{z}%
}$-direction. When they are anti-parallel $m_{y}^{K}\rightarrow -m_{y}^{K}$
in Eqs.~(\ref{exchange_couplings1}-\ref{exchange_couplings4}).

\subsection{Energy spectra of coupled NWA-spin wave modes}

With established interlayer dipolar and exchange coupling between the
lowest-branch spin waves in the film and acoustic mode in the nanowire
array, we can compute the energy spectra of the coupled system.

\subsubsection{Dominant interlayer dipolar coupling: anticrossings}

We first focus on the interlayer dipolar interaction, assuming that the
interlayer exchange interaction is efficiently suppressed by a thin spacer 
\cite{Co_YIG,CoFeB_YIG}. We then may use the approximate selection rule
found in Sec.~\ref{interlayer_dipolar_interaction}: when $\left\vert
k_{y}^{m}\right\vert \gtrsim s/\alpha_{\mathrm{ex}}$, the interlayer dipolar
coupling between the acoustic mode in the nanowire array and the
short-wavelength spin waves is chiral. This simplifies the analysis since
one only needs to consider the dipolar coupling between $\hat{\beta}_{%
\mathrm{H}k_{y}^{(m)}}$ and $\hat{\alpha}_{\mathrm{K}}$. For a particular $%
k_{y}^{(m)}$, the Hamiltonian of this subspace reads 
\begin{align}
\hat{H}(k_{y}^{(m)}) & =(1/2)(\hat{\beta}_{\mathrm{H}k_{y}^{(m)}}^{\dagger },%
\hat{\alpha}_{\mathrm{K}},\hat{\beta}_{\mathrm{H}k_{y}^{(m)}},\hat{\alpha }_{%
\mathrm{K}}^{\dagger})  \notag \\
& \times\left( 
\begin{array}{cccc}
\tilde{\omega}_{\mathrm{H}k_{y}^{(m)}} & \overline{B_{d}^{m}} & 0 & 
\overline{C_{d}^{m}} \\ 
B_{d}^{m} & \omega_{\mathrm{K}} & \overline{C_{d}^{m}} & 0 \\ 
0 & C_{d}^{m} & \tilde{\omega}_{\mathrm{H}k_{y}^{(m)}} & B_{d}^{m} \\ 
C_{d}^{m} & 0 & \overline{B_{d}^{m}} & \omega_{\mathrm{K}}%
\end{array}
\right) \left( 
\begin{array}{c}
\hat{\beta}_{\mathrm{H}k_{y}^{(m)}} \\ 
\hat{\alpha}_{\mathrm{K}}^{\dagger} \\ 
\hat{\beta}_{\mathrm{H}k_{y}^{(m)}}^{\dagger} \\ 
\hat{\alpha}_{\mathrm{K}}%
\end{array}
\right) ,  \label{blocks}
\end{align}
where $\tilde{\omega}_{\mathrm{H}k_{y}^{(m)}}$ and $\omega_{\mathrm{K}}$ are
the energies of the lowest-branch spin waves with momentum $k_{y}^{(m)}\hat{%
\mathbf{y}}$ and the NWA Kittel mode, respectively. When $\left\vert
B_{d}^{m}\right\vert \ll\tilde{\omega}_{\mathrm{H}k_{y}^{(m)}},\omega _{%
\mathrm{K}}$, terms with $B_d^m$ may be disregarded from rotating wave
approximation, the Hamiltonian is simplified to the quadratic form 
\begin{equation}
\hat{H}(k_{y}^{(m)})=(\hat{\beta}_{\mathrm{H}k_{y}^{(m)}}^{\dagger},\hat{%
\alpha}_{\mathrm{K}}^{\dagger})\left( 
\begin{array}{cc}
\tilde{\omega}_{\mathrm{H}k_{y}^{(m)}} & \overline{C_{d}^{m}} \\ 
C_{d}^{m} & \omega_{\mathrm{K}}%
\end{array}
\right) \left( 
\begin{array}{c}
\hat{\beta}_{\mathrm{H}k_{y}^{(m)}} \\ 
\hat{\alpha}_{\mathrm{K}}%
\end{array}
\right) ,
\end{equation}
with the frequencies 
\begin{equation}
\omega_{\pm}(k_{y}^{(m)})=\frac{\tilde{\omega}_{\mathrm{H}k_{y}^{(m)}}
+\omega_{\mathrm{K}}}{2}\pm\sqrt{\left( \frac{\tilde{\omega}_{\mathrm{H}
k_{y}^{(m)}}-\omega_{K}}{2}\right) ^{2}+|C_{d}^{m}|^{2}}.
\end{equation}
$|C_{d}^{m}|$ is the coupling strength between the short-wavelength spin
waves in the film and the acoustic mode in the nanowire array which governs
the anticrossing with splitting of $2|C_{d}^{m}|$ between these modes at the
resonance $\tilde{\omega}_{\mathrm{H}k_{y}^{(m)}}=\omega_{\mathrm{K}}$. In
Sec.~\ref{compare_with_exp}, we calculate this coupling strength for
experimental conditions in Refs.~\cite{Haiming_NC,Haiming_PRL}, which can be
used to understand the experiments \cite{Haiming_NC,Haiming_PRL} without
having to invoke interface exchange.

\subsubsection{Dominant interlayer exchange coupling: in-plane standing wave}

\label{in_plane_standing} When the interlayer exchange is active, we need to
additionally consider the couplings between $\hat{\alpha}_{\mathrm{K}}$, $%
\hat{\beta }_{\mathrm{H}k_{y}^{(m)}}$ and $\hat{\beta}_{\mathrm{H}%
-k_{y}^{(m)}}$. At resonance $\omega_{\mathrm{K}}=\omega_{\mathrm{H}\pm
k_{y}^{(m)}}\equiv \omega_{m}^{0}$, the Hamiltonian becomes 
\begin{equation}
\hat{H}\approx(\hat{\alpha}_{\mathrm{K}}^{\dagger},\hat{\beta}_{\mathrm{H}
k_{y}^{(m)}}^{\dagger},\hat{\beta}_{\mathrm{H}-k_{y}^{(m)}}^{\dagger})\left( 
\begin{array}{ccc}
\omega_{m}^{0} & C_{m} & D_{m} \\ 
\overline{C_{m}} & \omega_{m}^{0} & 0 \\ 
\overline{D_{m}} & 0 & \omega_{m}^{0}%
\end{array}
\right) \left( 
\begin{array}{c}
\hat{\alpha}_{\mathrm{K}} \\ 
\hat{\beta}_{\mathrm{H}k_{y}^{(m)}} \\ 
\hat{\beta}_{\mathrm{H}-k_{y}^{(m)}}%
\end{array}
\right) ,  \label{full_H}
\end{equation}
where $C_{m}=C_{d}^{m}+C_{\mathrm{ex}}^{m}$ and $D_{m}=D_{d}^{m}+D_{\mathrm{%
ex}}^{m}$. Its eigenvalues are 
\begin{align}
\omega_{1} & =\omega_{m}^{0},  \notag \\
\omega_{2} & =\omega_{m}^{0}-\sqrt{|C_{m}|^{2}+|D_{m}|^{2}},  \notag \\
\omega_{3} & =\omega_{m}^{0}+\sqrt{|C_{m}|^{2}+|D_{m}|^{2}},
\label{energies}
\end{align}
with corresponding eigenfunctions 
\begin{align}
\psi_{1} & =\left( 0,-D_{m}/C_{m},1\right) ,  \notag \\
\psi_{2} & =\left( -\sqrt{|C_{m}|^{2}+|D_{m}|^{2}}/D_{m},{C_{m}}/{D_{m}}%
,1\right) ,  \notag \\
\psi_{3} & =\left( \sqrt{|C_{m}|^{2}+|D_{m}|^{2}}/D_{m},{C_{m}}/{D_{m}}%
,1\right) .  \label{functions}
\end{align}

When the interlayer exchange interaction is much larger than the dipolar
one, $C_{m}\approx D_{m}$. In this situation, the first eigenfunction in
Eq.~(\ref{functions}) corresponds to the \textit{in-plane standing wave} in
the film, which arises from the linear superposition of the two spin waves
with opposite momenta.

\section{Material and device parameter dependence}

\label{compare_with_exp}

In this section, we illustrate the expressions we produced above and
demonstrate the magnitude of the effect by specifically considering coupling
between a nanowire array and a thin film for the Co or Ni NWAs fabricated on
YIG films. This system has been experimentally realized in Refs. \cite%
{Haiming_NC,Haiming_PRL}, and we use the parameters from these papers.

\subsection{Co nanowire array}

The lattice constant of the Co nanowire array was $a=180$~nm with wire
thickness $h=20$~nm and width of $d=132$~nm \cite{Haiming_NC,Haiming_PRL}.
The saturated magnetization $\mu_{0}{M}_{0}=1.1$~T for the Co and $\mu_{0}%
\tilde{M}_{0}=0.177$~T for the YIG films. The YIG exchange interaction
constant is $\alpha_{\mathrm{ex}}=3\times10^{-16}$~m$^{2}$ \cite%
{exchange_stiffness} and thickness $s=20$~nm \cite{Haiming_PRL}. We compute
the coupling constants for these parameters when the magnetizations in the
nanowire array and film are parallel and antiparallel to each other in Figs.~%
\ref{Co}(a) and (b), respectively, as a function of the mode index $m={\pi}%
/\left( k_{y}^{(m)}{a}\right) {\ }${of} allowed spin waves in YIG. The
magnetic field is chose to be constant and to agree with a main
anticrossing. In Fig.~\ref{Co}(a), for example, $\mu_{0}H_{z}=0.012$~T
corresponds to the anticrossing of the mode $m=4.$

\subsubsection{Parallel configuration}

\begin{widetext}
\begin{figure}[tbh]
\begin{minipage}[]{18cm}
			\hspace{0 cm}\parbox[t]{8.5cm}{
				\includegraphics[width=8.2cm]{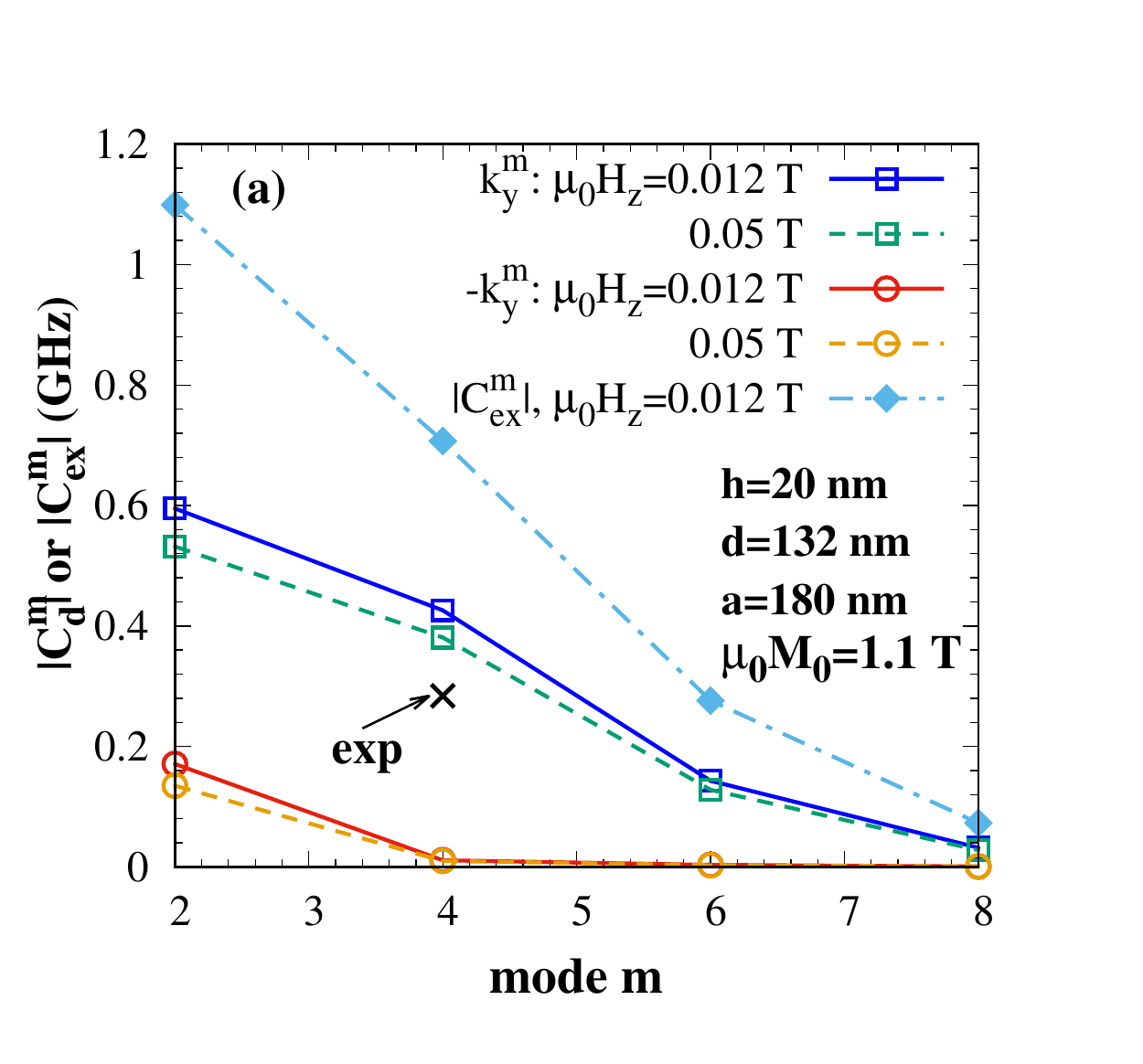}}
			\hspace{0cm}\parbox[t]{8.5cm}{
				\includegraphics[width=8.2cm]{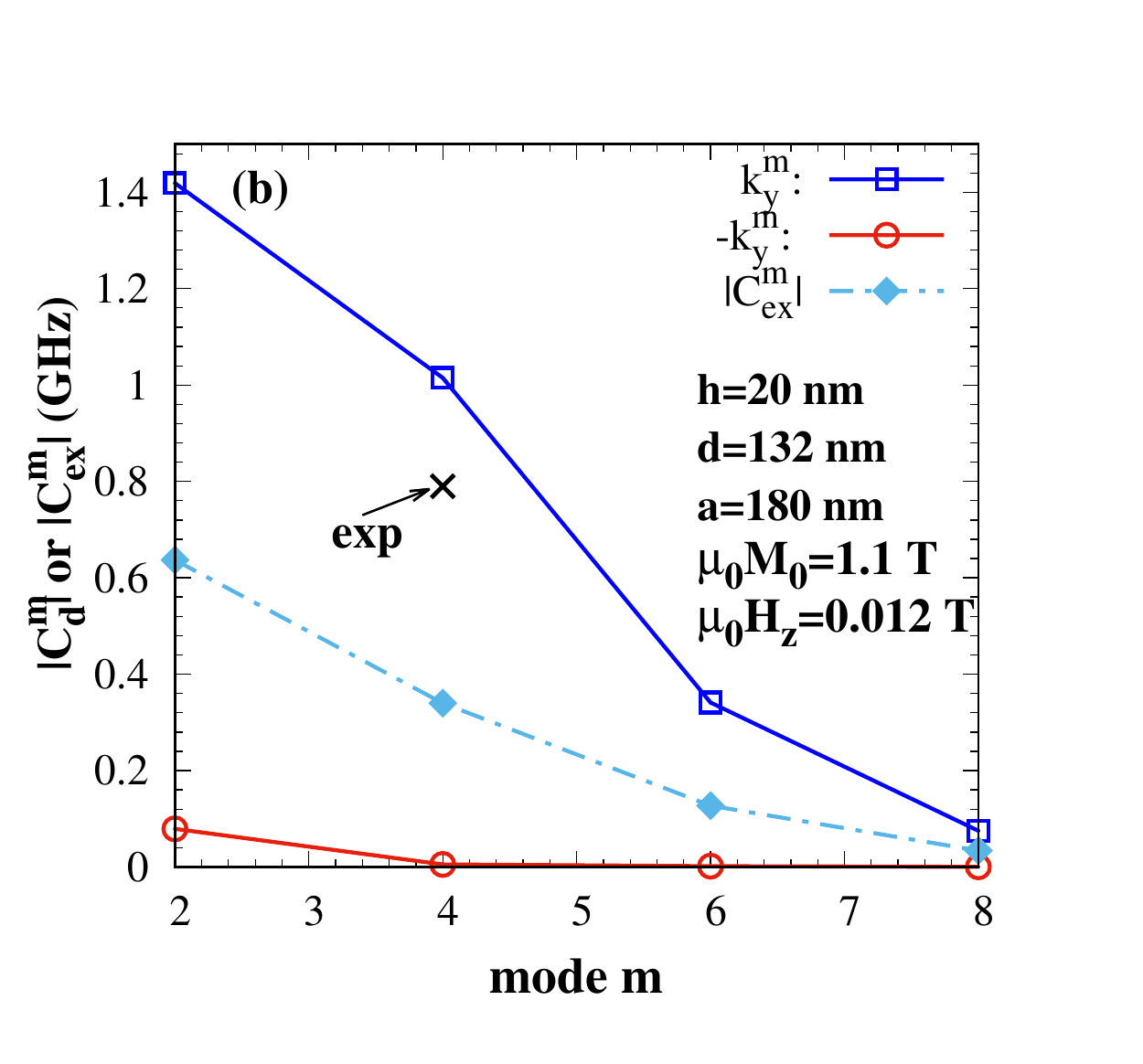}}
		\end{minipage} 
\begin{minipage}[]{18cm}
			\begin{center}
				\caption{(Color online) Mode
					dependence of the interlayer dipolar and exchange couplings between Co
					nanowires and a YIG film for parallel (a) and antiparallel (b) magnetizations.
					In (a) the blue (red) solid curve with squares (circles) represents the
					interlayer dipolar coupling between the spin waves with momentum $k_{y}%
					^{m}\hat{\mathbf{y}}$ ($-k_{y}^{(m)}\hat{\mathbf{y}}$) in the film and the
					Kittel mode mode of the NWA for $\mu_{0}H_{z}=0.05$~T, while the cyan
					dot-dashed curve with diamonds denotes the interlayer exchange coupling (which
					is the same for spin wave directions). Analogous curves are plotted for
					$\mu_{0}H_{z}=0.05$~T (the exchange contribution does not depend on the
					field), The crosses in (a) and (b) denote the anticrossing gaps observed in
					FMR experiments \cite{Haiming_PRL}. $m$ is an even integer.}%
				\label{Co}
			\end{center}
		\end{minipage}
\end{figure}
\end{widetext}

When $\mathbf{M}_{0}\parallel\mathbf{\tilde{m}}_{0}\parallel\mathbf{H}_{%
\mathrm{app}}\parallel\hat{\mathbf{z}}$ the mode dependence of the dipolar
coupling strengths is shown in Fig.~\ref{Co}(a) with applied magnetic fields 
$\mu_{0}H_{z}=0.012$ and 0.05~T. When $\mu_{0}H_{z}=0.012$~T, in Fig.~\ref%
{Co}(a), the blue (red) solid curve with squares (circles) describes the
mode dependence of the interlayer dipolar coupling between the lowest spin
wave subband with momentum $k_{y}^{(m)}\hat{\mathbf{y}}$ ($-k_{y}^{(m)}\hat{%
\mathbf{y}}$) in the YIG film and the FMR of the Co nanowire array. The
coupling strength for the spin waves with positive wave vector $k_{y}^{(m)}%
\hat{\mathbf{y}}$ is much larger than that for opposite $-k_{y}^{(m)}\hat{%
\mathbf{y}}$ one when $m\ge 4$ that corresponds to exchange spin waves,
confirming that the chirality of the coupling should be very significant in
real systems.

With increasing mode number, the coupling strength decreases. According to
Eq.~(\ref{dipolar_couplings}), $C_{d}^{m}\propto
F_{m}\int_{-s}^{0}dxe^{k_{y}^{(m)}x}$, where $F_{m}$ is the Fourier
component of the NWA magnetization dynamics, while the integral represents
the decay of the dipolar field inside the film. The drop of the coupling
with increasing $m$ is caused by the evanescent decay of the dipolar field
and not by the form factor $F_{m}\propto\sin(k_{y}^{(m)}d/2)$. In the
presence of a non-magnetic insertion with thickness $\delta$, the overlap
integral 
\begin{equation}
\int_{-s-\delta}^{-\delta}dxe^{k_{y}^{(m)}x}=\int_{-s}^{0}dxe^{k_{y}
^{(m)}(x-\delta)}=e^{-k_{y}^{m}\delta}\int_{-s}^{0}dxe^{k_{y}^{(m)}x}.
\end{equation}
So the inserted layer exponentially suppresses the interlayer dipolar
coupling by $e^{-k_{y}^{m}\delta}$. However, this effect is rather
inefficient for $\delta=1~\mathrm{nm}$ and a wave length $2\pi/k_{y}^{m}=100$%
~nm, i.e. $k_{y}^{m}\delta=\pi/50\approx0.06$.

The decrease of the coupling with magnetic field in Fig.~\ref{Co}(a) can be
understood as follows. For relatively short-wavelength spin waves with $%
\tilde{m}_{0,y}^{k_{y}^{(m)}}\approx i\tilde{m}_{0,x}^{k_{y}^{(m)}}$, Eq.~(%
\ref{dipolar_couplings}) gives 
\begin{equation}
C_{d}^{m}\approx-2\mu_{0}\gamma\sqrt{\tilde{M}_{0}M_{0}}\frac{F_{m}}{k_{y}
^{m}}(1-e^{-k_{y}^{(m)}s})\tilde{m}_{0,x}^{k_{y}^{(m)}}(m_{x}^{\mathrm{K}}-i%
\overline{m_{y}^{\mathrm{K}}}).
\end{equation}
For $s/\alpha_{\mathrm{ex}}\lesssim\left\vert k_{y}\right\vert \lesssim1/s$
the amplitudes $\tilde{m}_{0,x}^{k_{y}^{(m)}}$ and $\tilde{m}%
_{0,y}^{k_{y}^{(m)}}$ in the film do not depend strongly on the field, in
contrast to the NWA Kittel mode. Specifically, 
\begin{equation}
-(m_{x}^{\mathrm{K}}-i\overline{m_{y}^{\mathrm{K}}})=(\mathcal{F}^{1/4}-%
\mathcal{F}^{-1/4})\sqrt{a/(4hd)},  \label{suppression}
\end{equation}
in which $\mathcal{F}\equiv(H_{\mathrm{app}}^{z}+M_{0}^{z}N_{xx})/(H_{%
\mathrm{app}}^{z}+M_{0}^{z}N_{yy})$. When $N_{xx}\gg N_{yy}$ and $%
M_{0}^{z}\gg H_{\mathrm{app}}^{z}$, $\mathcal{F}\approx M_{0}^{z}N_{xx}/(H_{%
\mathrm{app}}^{z}+M_{0}^{z}N_{yy})$ decreases with $H_{\mathrm{app}}^{z}$
and so does the interlayer dipolar coupling.

We also present the interlayer exchange coupling $C_{ex}^{\left( m\right) }$
for direct contact between the Co NWA and the YIG film by the cyan
dot-dashed curve with diamonds in Fig.~\ref{Co}(a), with an interlayer
exchange coupling constant $J=200$~$\mathrm{\mu}$J/m$^{2}$ \cite{Co_YIG}.
Without spacer layer, the interlayer exchange coupling wins over the dipolar
interaction for the sample geometries considered here. The decrease can be
understood from $C_{\mathrm{ex}}^{m}$ in Eq.~(\ref{exchange_couplings2}): $%
\tilde{m}_{0,x}^{k_{y}^{(m)}}$, $\tilde{m}_{0,y}^{k_{y}^{(m)}}$, $m_{x}^{%
\mathrm{K}}$ and $m_{y}^{\mathrm{K}}$ do not depend strongly on mode number,
but we find a decreasing $\left\vert C_{\mathrm{ex}}^{m}\right\vert
\propto\left\vert \sin(k_{y}^{(m)}d/2)/m\right\vert $ with increasing $m$. $%
C_{\mathrm{ex}}^{m}$ can also become oscillatory as a function of $m$ (refer
to Sec.~\ref{Ni_numerical} below).

\subsubsection{Antiparallel configuration}

Assuming that $\mathbf{\tilde{m}}_{0}\parallel \mathbf{H}_{\mathrm{app}%
}\parallel \hat{\mathbf{z}}$ and $\mathbf{M}_{0}\parallel -\hat{\mathbf{z}}$%
, $M_{0}^{z}$ becomes negative and $m_{y}^{K}$ is replaced by $-m_{y}^{K}$
when calculating the interlayer dipolar and exchange couplings. The results
in Fig.~\ref{Co}(b) for $\mu _{0}H_{z}=0.012$~T show a strong enhancement of
the magnitude and chirality of the dipolar coupling at the cost of a reduced
exchange interaction, which is caused by$\left\vert m_{x}^{K}-i\overline{%
m_{y}^{K}}\right\vert <\left\vert m_{x}^{K}+i\overline{m_{y}^{K}}\right\vert 
$, see Eq.~(\ref{wave_nanowire}). However, we disregarded here a possible
exchange-spring magnetization texture in the non-collinear and antiparallel
configurations \cite{Haiming_PRL}, whose treatment is beyond of the scope of
this work.

\subsection{Ni nanowire array}

\label{Ni_numerical}

\begin{figure}[th]
\begin{center}
{\includegraphics[width=8.2cm]{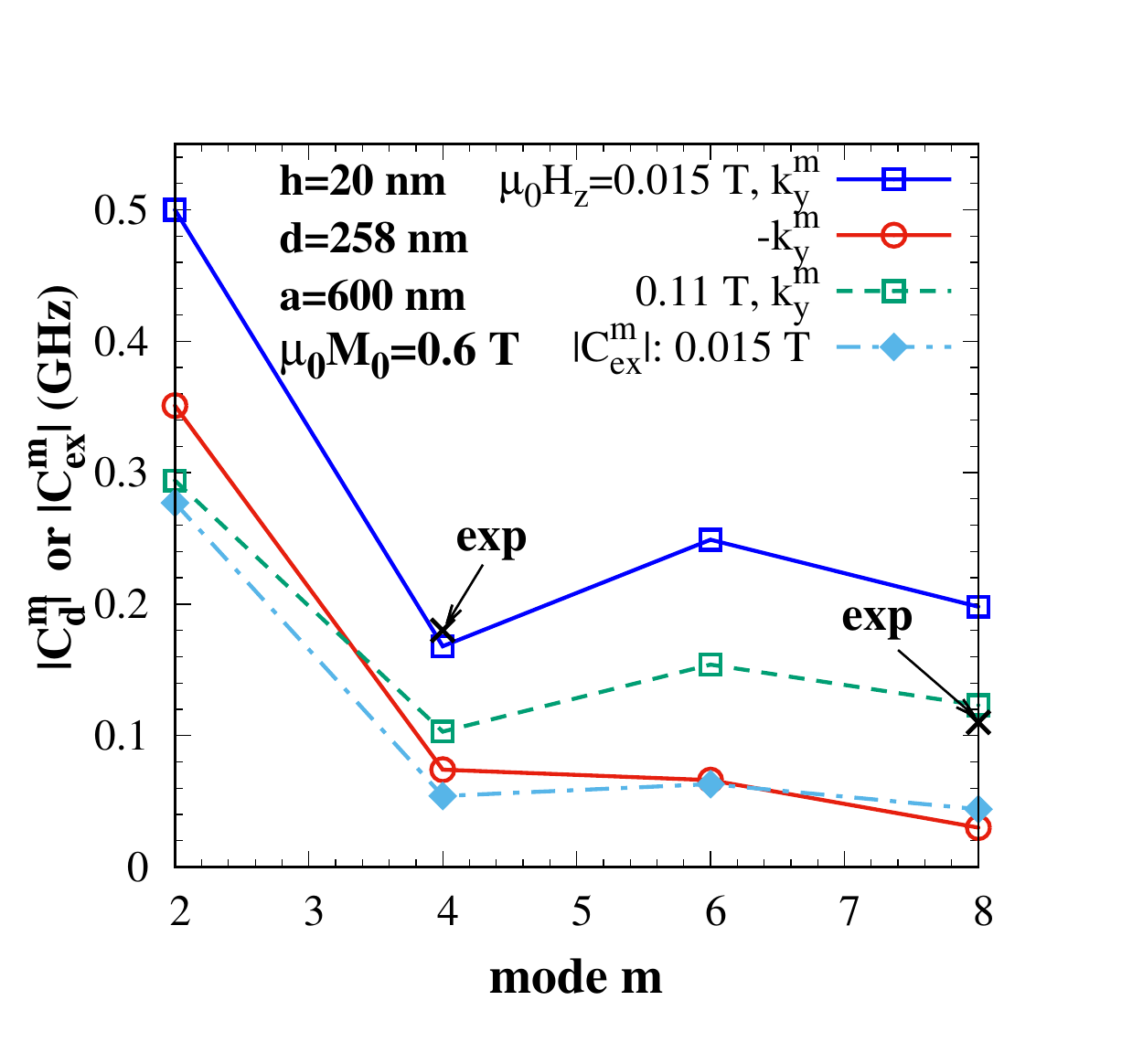}}
\end{center}
\caption{(Color online) Mode dependences of the interlayer dipolar and
exchange couplings betweeen a Ni NWA and YIG film when magnetizations and
applied field are all parallel along the wires. The blue solid curve with
squares and red solid curve with circles represent the interlayer dipolar
couplings for the spin waves with momenta $k_{y}^{(m)}\hat{\mathbf{y}}$ and $%
-k_{y}^{(m)}\hat{\mathbf{y}}$, respectively, for $\protect\mu %
_{0}H_{z}=0.015 $~T. The green dashed curve with squares is the interlayer
dipolar coupling for positive momenta and $\protect\mu _{0}H_{z}=0.11$~T.
The cyan dot-dashed curve with diamonds denotes the interlayer exchange
coupling for momenta $\pm k_{y}^{(m)}\hat{\mathbf{y}}$ when $\protect\mu %
_{0}H_{z}=0.015$~T. The crosses are the mode splittings observed in the FMR 
\protect\cite{Haiming_PRL}.}
\label{Ni}
\end{figure}

Experiments have been also carried out on a Ni NWA with a (relatively large)
lattice constant $a=600$~nm, and thickness and width of $h=20$~nm and $d=258$%
~nm, respectively, and with a thin spacer of $1$~nm between Ni wires and the
YIG substrate \cite{Haiming_NC,Haiming_PRL}. The Ni saturated magnetization
is $\mu _{0}{M}_{0}=0.6$~T \cite{Haiming_PRL}. For these parameter the
factor $\sim \sin (k_{y}^{(m)}d/2)$ causes a non-monotonous dependence of
the interlayer dipolar coupling, see Fig.~\ref{Ni}. For $\mu _{0}H_{z}=0.015$%
~T, the asymmetry in the coupling of the Kittel mode to spin waves
propagating into opposite directions is strong, for $m\geq 4$, the chirality
is almost perfect. For larger $\mu _{0}H_{z}=0.11$~T the interlayer dipolar
coupling is suppressed for the same reason as for the Co NWA discussed above.

The interlayer exchange coupling is also shown in Fig.~\ref{Ni} for an
exchange interaction strength $J=30~\mathrm{\mu }$J/m$^{2}$ \cite%
{Haiming_PRL}, which is smaller than the dipolar one.


\subsection{Summary of the comparison with experiments}

The present study was motivated by FMR experiments which displayed clear
anti-crossings, i.e. strong coupling, between YIG film and NWA spin wave
modes \cite{Haiming_NC,Haiming_PRL}. The observed splittings are shown by
the crosses in Figs.~\ref{Co} and \ref{Ni}, respectively. The experimental
values are quite close to the calculated ones for dipolar interactions
without fit parameters. This supports the assumption that interlayer
exchange interactions are suppressed by spacer layers inserted between the
YIG film and Co/Ni nanowires \cite{CoFeB_YIG,Co_YIG,Haiming_NC,Haiming_PRL}.
A dominant interlayer dipolar interaction implies a chiral coupling. As
shown in Figs.~\ref{Co} and \ref{Ni}, only the short-wavelength spin waves
propagating with momenta $\mathbf{k}\parallel \mathbf{\tilde{m}}\times 
\mathbf{n\ }$interact with the NWA Kittel mode, where $\mathbf{n}$ is the
unit vector normal to the interface. A similar chiral feature is intrinsic
to the Damon-Eshbach surface mode that exist in sufficiently thick films 
\cite{DE} but not in the ultrathin films considered here.

\section{Conclusion and discussion}

\label{summary} In conclusion, we demonstrated that spin waves can be
coherently excited in an ultrathin magnetic film in only one direction by a
magnetic grating. We focus on the limiting cases in which the applied
magnetic field and magnetizations in the film are either parallel or
antiparallel to the NWA magnetization and wire axis. We report an unexpected
chirality in the coupling that strongly favors spin waves propagating
perpendicular to the nanowires with wave vector$\ \left\vert \mathbf{k}%
\right\vert =m\pi /a$ (where $m$ is an even integer and $a$ the NWA\ lattice
constant) \cite{Haiming_NC,Haiming_PRL}. The dipolar regime can be realized
by an inserted non-magnetic layer between the YIG film and nanowires that
suppresses the exchange interaction more efficiently than the dipolar one 
\cite{CoFeB_YIG,Co_YIG,Haiming_PRL}. The calculated coupling strength agrees
well with the experimental observations \cite{Haiming_PRL} for both parallel
and anti-parallel configurations. This suggests that the interlayer dipolar
interaction plays a dominant role in the experiments \cite{Haiming_PRL}, but
more work, especially including magnetic texture and dynamic exchange is
necessary to confirm this assertion. The spacer layer might also be
instrumental to support an antiparallel magnetic configuration without
associated exchange-spring magnetization textures in the film.

The dipolar coupling is a classical interaction
between two magnetic bodies that has a relative longer range than the
(static) exchange interaction. In the present configuration both
interactions are exponentially suppressed with distance between the magnets,
but on an atomic length scale and that of the wave length for exchange and
dipolar coupling. In ultra-thin films without chiral surface waves, the
exchange coupling mixes the Kittel mode almost symmetrically with the spin
waves in opposite directions, thereby leading to in-plane \textit{standing}
waves by interference. In the presence of spacer layers, the dynamic
exchange interaction competes as well, falling off on the scale of the
spin-flip diffusion length, which can be rather long-range when the spacer
is a clean simple metal such as copper \cite{Klingler12}.

The spin waves with the wave vector $\mathbf{k}$ in a thin film with surface normal $\mathbf{n}$
are coherently exited by the NWA grating with equilibrium magnetization
along $\mathbf{\tilde{m}}$ and propagate dominantly in the direction $\mathbf{k}%
\parallel (\mathbf{\tilde{m}}\times \mathbf{n})$ (but only for significantly
elliptic precession of either NWA or film magnetic modes). This
phenomenology agrees with the intrinsic chirality of dipolar Damon-Eshbach
surface modes in thick films \cite{DE}. However, the physics which we
describe in this work is quite different, since there is no intrinsic
chirality in the spin waves of ultrathin magnetic films with nearly constant
amplitude over the film thickness. It is rather the intrinsic chirality of
the dipolar fields that generates a chiral coupling to non-chiral spin
waves. This directionality can be exploited in several ways \cite%
{in_preparation}, for example, to generate a heat conveyer belt \cite%
{heatconveyer1,heatconveyer2,heatconveyer3,heatconveyer4} without the need
for surface states.

Finally, we would point out an electric analogy, viz. the chiral coupling
induced by rotating electric (rather than magnetic) dipoles. When excited
close to a planar waveguide, the chiral evanescent electromagnetic field
unidirectionally excites surface plasmon polaritons \cite{science}, also
referred to as \textquotedblleft spin-orbit interaction of
light\textquotedblright\ \cite{Petersen}. There are large differences in the
physics that we will emphasize elsewhere, but note that the dipolar field
with momentum larger than $\omega /c$ with $\omega $ and $c$ being the
frequency and light velocity is evanescent on a sub-wavelength scale. Its
chirality arises from the near-field interference of the radiated fields
from the vertical and horizontal components of the ac electric field \cite%
{science}. The circularly-polarized magnetic dipolar dynamics generates a
purely circularly-polarized magnetic field [e.g., see Eq.~(\ref%
{circular_field})], while the circularly-polarized electric dipole results
in an elliptically-polarized field by retardation [see Eq.~(1) in Ref.~\cite%
{science}]. Nonetheless of this and other differences, the application
perspective of the chiral coupling found in plasmonics such as broadband
optical nanorouting \cite{science,Petersen} and polarization analyzers
 \cite{Espinosa} should stimulate similar activities in magnonics.

\begin{acknowledgments}
This work is financially supported by the Nederlandse Organisatie voor Wetenschappelijk Onderzoek (NWO) as well as JSPS KAKENHI Grant No. 26103006. One of the authors (TY) would like to thank Sanchar Sharma and Jilei Chen for useful discussions.
\end{acknowledgments}

\appendix

\section{Green function tensor}

\label{BB} Here we review the calculation of the demagnetizing field \cite%
{Landau} 
\begin{equation}
\tilde{\mathbf{H}}_{\beta }^{D}=\frac{1}{4\pi }\partial _{\beta }\int \frac{%
\partial _{\alpha }\tilde{\mathbf{M}}_{\alpha }(\mathbf{r}^{\prime })}{|%
\mathbf{r}-\mathbf{r^{\prime }}|}d\mathbf{r}^{\prime }.
\end{equation}%
in a thin magnetic film \cite{Kalinikos}. For a plane wave modulation $%
\tilde{\mathbf{M}}_{\alpha }(\mathbf{r})=\tilde{\mathbf{m}}_{\alpha
}(x)e^{ik_{y}y}e^{ik_{z}z}$, 
\begin{align}
\tilde{\mathbf{H}}_{\beta }^{D}& =\frac{1}{2}\partial _{\beta }\left\{
\int_{-s}^{0}dx^{\prime }\left[ -\tilde{\mathbf{m}}_{x}(x^{\prime })\mathrm{%
sgn}(x-x^{\prime })+\tilde{\mathbf{m}}_{y}(x^{\prime })\frac{ik_{y}}{|%
\mathbf{k}_{\parallel }|}\right. \right.  \notag \\
& \left. \left. +\tilde{\mathbf{m}}_{z}(x^{\prime })\frac{ik_{z}}{|\mathbf{k}%
_{\parallel }|}\right] e^{-\mathrm{sgn}(x-x^{\prime })(x-x^{\prime })|%
\mathbf{k}_{\parallel }|}e^{i\mathbf{k}_{\parallel }\cdot \mathbf{r}%
_{\parallel }}\right\} .
\end{align}%
In matrix form 
\begin{equation}
\left( 
\begin{array}{c}
\tilde{\mathbf{H}}_{x}^{D}(\mathbf{r}) \\ 
\tilde{\mathbf{H}}_{y}^{D}(\mathbf{r}) \\ 
\tilde{\mathbf{H}}_{z}^{D}(\mathbf{r})%
\end{array}%
\right) =e^{i\mathbf{k}_{\parallel }\cdot \mathbf{r}_{\parallel
}}\int_{-s}^{0}dx^{\prime }\mathcal{G}(x-x^{\prime })\left( 
\begin{array}{c}
\tilde{\mathbf{m}}_{x}(x^{\prime }) \\ 
\tilde{\mathbf{m}}_{y}(x^{\prime }) \\ 
\tilde{\mathbf{m}}_{z}(x^{\prime })%
\end{array}%
\right) ,  \label{Green}
\end{equation}%
where 
\begin{widetext}
\begin{equation}
\mathcal{G}(x-x^{\prime })=e^{-|x-x^{\prime }||\mathbf{k}_{\parallel
}|}\left( 
\begin{array}{ccc}
\frac{|\mathbf{k}_{\parallel }|}{2} & -\frac{ik_{y}}{2}\mathrm{sgn}\left(
x-x^{\prime }\right) & -\frac{ik_{z}}{2}\mathrm{sgn}\left( x-x^{\prime
}\right) \\ 
-\frac{ik_{y}}{2}\mathrm{sgn}\left( x-x^{\prime }\right) & -\frac{k_{y}^{2}}{%
2|\mathbf{k}_{\parallel }|} & -\frac{k_{y}k_{z}}{2|\mathbf{k}_{\parallel }|}
\\ 
-\frac{ik_{z}}{2}\mathrm{sgn}\left( x-x^{\prime }\right) & -\frac{k_{y}k_{z}%
}{2|\mathbf{k}_{\parallel }|} & -\frac{k_{z}^{2}}{2|\mathbf{k}_{\parallel }|}%
\end{array}%
\right) -\delta _{\mathrm{in}}(x-x^{\prime })\mathcal{I}
\end{equation}%
\end{widetext}
is the Green function and $\mathcal{I}$ the unity tensor. The $\delta _{%
\mathrm{in}}$-function vanishes when $x$ lies outside the magnetic film. The
demagnetization field $\tilde{\mathbf{H}}^{D}$ naturally satisfies
electromagnetic boundary condition, i.e. continuity of the electromagnetic
fields and currents at the surface of the magnet \cite{Kalinikos}.

\section{Higher magnon subbands in thin films}

\label{CC} Here we estimate the effects of higher-order standing wave modes
on the spin waves in the lowest subband. Retaining only the lowest-order
modes in Eq.~(\ref{eigen_full}) we arrive at the secular equation 
\begin{widetext}
{\scriptsize \ 
\begin{equation}
\left( 
\begin{array}{cccc}
\tilde{\omega}+\Omega_{H}+\alpha_{\mathrm{ex}}k_{y}^{2}+\frac{1}{2} & \frac {%
1}{2}-\frac{1}{2}|k_{y}|Q_{00} & 0 & -\frac{k_{y}}{2}\tilde{Q}_{01} -\frac{%
|k_{y}|}{2}Q_{01} \\ 
-\frac{1}{2}+\frac{1}{2}|k_{y}|Q_{00} & \tilde{\omega}-\Omega_{H} -\alpha_{%
\mathrm{ex}}k_{y}^{2}-\frac{1}{2} & -\frac{k_{y}}{2}\tilde{Q} _{01}+\frac{%
|k_{y}|}{2}Q_{01} & 0 \\ 
0 & -\frac{k_{y}}{2}\tilde{Q}_{10}-\frac{|k_{y}|}{2}Q_{10} & \tilde{\omega }%
+\Omega_{H}+\alpha_{\mathrm{ex}}k_{y}^{2}+\alpha_{\mathrm{ex}}(\frac{\pi} {s}%
)^{2}+\frac{1}{2} & \frac{1}{2}-\frac{1}{2}|k_{y}|Q_{11} \\ 
-\frac{k_{y}}{2}\tilde{Q}_{10}+\frac{|k_{y}|}{2}Q_{10} & 0 & -\frac{1} {2}+%
\frac{1}{2}|k_{y}|Q_{11} & \tilde{\omega}-\Omega_{H}-\alpha_{\mathrm{ex}
}k_{y}^{2}-\alpha_{\mathrm{ex}}(\frac{\pi}{s})^{2}-\frac{1}{2}%
\end{array}
\right) \left( 
\begin{array}{c}
\tilde{{m}}_{+,0} \\ 
\tilde{{m}}_{-,0} \\ 
\tilde{{m}}_{+,1} \\ 
\tilde{{m}}_{-,1}%
\end{array}
\right) =0. 
 \label{higher_order}
\end{equation}
} 
\end{widetext}
With Eq.~(\ref{QQs}) we find $Q_{01}=0$ and $\tilde{Q}_{01}={2\sqrt{2} s}%
(e^{-|k_{y}|s}+1)/(k_{y}^{2}s^{2}+\pi^{2})$. The matrix in Eq.~(\ref%
{higher_order}) can be directly diagonalized by the Bogoliubov
transformation \cite{Kostylev}. We can use perturbation theory to estimate
the importance of higher-order modes for our film thicknesses. The second
mode contributes to the with amplitudes $c_{k}\equiv k_{y}\tilde{Q}%
_{01}/[2\alpha_{\mathrm{ex}}(\pi/s)^{2}]$. For a grating with period $a=180$%
~nm \cite{Haiming_NC,Haiming_PRL}, $k_{y}=4\pi/a$, while thickness of the
film is $s=20$~nm. Then $k_{y}\tilde{Q}_{01}/2=0.208$, $\alpha_{\mathrm{ex}%
}(\pi/s)^{2}\approx7.40$, and hence $c_{k}\approx0.028$, which can be safely
disregarded.

\end{document}